\documentclass{aa}
\usepackage[normalem]{ulem}
\usepackage{natbib,twoopt}
\usepackage[colorlinks=true, allcolors=blue]{hyperref}
\usepackage{threeparttable}
\usepackage{booktabs}
\usepackage{amsmath}
\usepackage{graphicx}
\usepackage{txfonts}
\usepackage{color}
\usepackage{multirow}
\usepackage[T1]{fontenc}

\begin{document}

\title{Uncertainties of the dust grain size in protoplanetary disks retrieved from millimeter continuum observations}
    \author{Dafa Li\inst{1,2} 
	        \and 
			Yao Liu\inst{1}
			\and
			Hongchi Wang\inst{1}
			\and
			Min Fang\inst{1}
			\and
			Lei Wang\inst{1}
			}
    \institute{Purple Mountain Observatory \& Key Laboratory of Radio Astronomy, Chinese Academy of Sciences, 10 Yuanhua Road, Qixia District, Nanjing 210023, China\\
    \email{yliu@pmo.ac.cn}
         \and
    School of Astronomy and Space Science, University of Science and Technology of China, 96 Jinzhai Road, Hefei 230026, P. R. China\\
    }
 
   \abstract
   {Investigating the dust grain size and its dependence on substructures in protoplanetary disks is a crucial step in understanding the initial process of planet formation. Spectral indices derived from millimeter observations are used as a common probe for grain size. Converting observed spectral indices into grain sizes is a complex task that involves solving the radiative transfer equation, taking into account the disk structure and dust properties.}
   {Under the assumption of vertically isothermal disks, the solution to the radiative transfer equation can be approximated with an analytic expression, with which the fitting procedure can be done very fast. Our work aims to investigate the applicability of this method to grain size retrieval.}
   {We ran reference radiative transfer models with known disk properties, and generated four synthetic images at wavelengths of 0.8, 1.3, 3, and 7.8\,mm, representing high-resolution continuum observations. Rings and gaps were considered in the setup. We fit the synthetic images using the analytic solution to investigate the circumstances under which the input grain sizes can be recovered.}
   {Fitting images at only two wavelengths is not sufficient to retrieve the grain size. Fitting three images improves the retrieval of grain size, but the dust surface density is still not well recovered. When taking all of the four images into account, degeneracies between different parameters are highly reduced, and consequently the best-fit grain sizes are consistent with the reference setup at almost all radii. We find that the inclination angle has a significant impact on the fitting results. For disks with low inclinations, the analytic approach works quite well. However, when the disk is tilted above ${\sim}\,60^{\circ}$, neither the grain size nor the dust surface density can be constrained, as the inclination effect will smooth out all substructures in the radial intensity profile of the disk.}
   {}

   \maketitle

\section{Introduction}
\label{sect:intro}
Dust grains in protoplanetary disks are the building blocks of planets. During the planet formation process, the size of dust particles increases by more than ten orders of magnitude \citep[e.g.,][]{Armitage2010}. It is generally believed that micron-sized dust grains collide and stick together by van der Waals forces, forming large aggregates up to the scale of millimeters, and that the gravitational force between kilometer-sized planetesimals is enough to accrete surrounding dust, making the bodies grow to larger sizes \citep[e.g.,][]{Chokshi1993,Dominik1997,Birnstiel2016,Drazkowska2023,Birnstiel2023}. 

In existing theories, it is difficult for dust grains to grow from the scale of millimeters to that of meters; this is known as the meter-size barrier \citep{1977MNRAS.180...57W}.  On the millimeter scale, the relative velocities between dust particles become significant, causing them to collide with higher energies. The increased energy in collisions often leads to bouncing instead of sticking, preventing the grains from further growth \citep[e.g.,][]{Testi2014}. Furthermore, radial drift is another obstacle to grain growth. Dust grains lose angular momentum due to the sub-Keplerian motion of gas in protoplanetary disks, and so fast migrate toward the central star \citep{1972fpp..conf..211W,1976PThPh..56.1756A,1986Icar...67..375N,Youdin2002,Brauer2007}. Consequently, dust grains, especially millimeter pebbles, will drift to the inner region before they can grow into larger sizes \citep{Takeuchi2002,2008A&A...480..859B,2009A&A...503L...5B}. However, mechanisms that facilitate dust grains to accumulate and grow likely work at specific locations in the disk. 

High-resolution Atacama Large Millimeter/submillimeter Array (ALMA) observations reveal that substructures, for instance rings and gaps, are common in protoplanetary disks \citep[e.g.,][]{Long2018,Andrews2018,Andrews2020,Bae2023}. It has been proposed that the origin of these finely scaled features is related to various processes, such as (magneto-)hydrodynamic instabilities \citep[e.g.,][]{Lesur2022}, snow lines \citep[e.g.,][]{Zhang2015}, and planets \citep[e.g.,][]{Kley2012,Paardekooper2022}. From a theoretical perspective, disk substructures act as pressure bumps that can trap dust particles, which increases the density and facilitates grain growth \citep{1972fpp..conf..211W,2012A&A...538A.114P,2018ApJ...860L..12T,2018ApJ...868..113T,2020MNRAS.495..173R}. Investigating the relationship between grain size and 
substructures is crucial to understanding dust evolution and developing models of planet formation. 

\begin{figure}[!t]
\centering
\includegraphics[width=\columnwidth,angle=0]{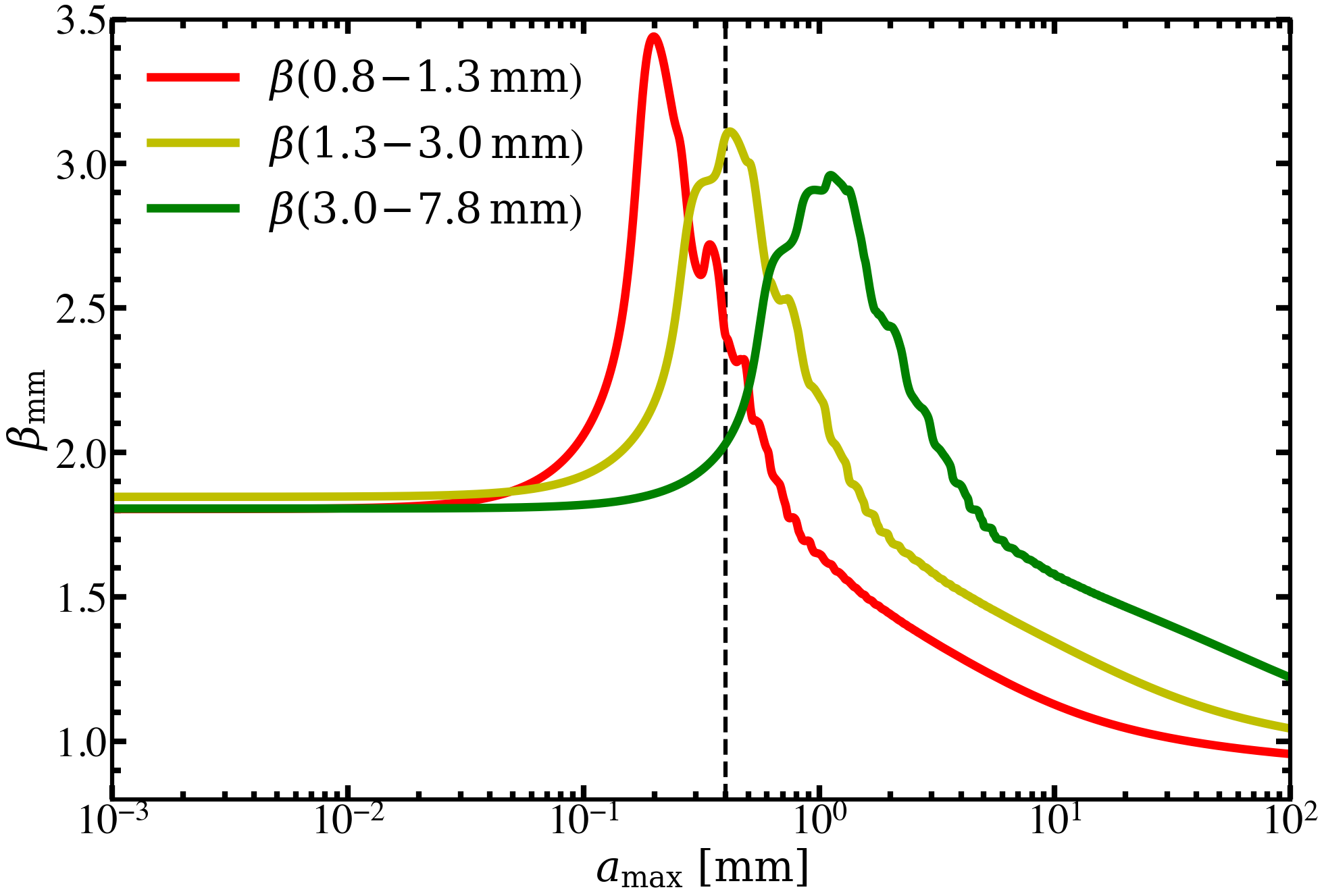}
\caption{Dust opacity slope, $\beta_{\mathrm {mm}}$, as a function of the maximum grain size, $a_{\rm max}$. The slopes between two wavelengths given in the legend are derived by calculating the mass absorption coefficients ($\kappa_{\rm abs}$). To calculate $\kappa_{\rm abs}$, we adopted the MIE scattering theory, assuming the DSHARP dust composition (see Sect.~\ref{section:dust}). The grain size distribution follows a power law, ${N(a)}\,{\propto}\,a^{-3.5}$, with the minimum dust size fixed to $a_{\mathrm{min}}\,{=}\,0.01\,\mu \mathrm{m}$. The vertical dashed line marks the 0.4\,mm boundary that divides the $a_{\rm max}$ values into two regimes (a small size region and a large size region) according to the $\beta(1.3-3\,\rm{mm})$ curve.}
\label{fig:beta}
\end{figure}

A common probe for grain size is the millimeter spectral index $\alpha_{\rm {mm}}\,{=}\,\log \left(I_{\nu_1} / I_{\nu_2}\right) / \log \left(\nu_1 / \nu_2\right)$, where $I_{\nu_1}$ and $I_{\nu_2}$ are the measured flux densities at frequencies $\nu_1$ and $\nu_2$, respectively. In the (sub)millimeter regime, the mass absorption coefficient ($\kappa_{\nu}$) of dust grains can be approximated as $\kappa_{\nu}\,{\propto}\,\nu^{\beta_{\mathrm {mm}}}$. In the optically thin case, the spectral index is linked to the opacity slope via $\beta_{\mathrm {mm}}\,{=}\,\alpha_{\rm mm}\,{-}\,\log (B_{\nu 1} / B_{\nu 2}) / \log (\nu_{1} / \nu_{2})$, where $B_{\nu}$ is the Planck function. With a given dust composition and shape, the opacity slope, $\beta_{\mathrm {mm}}$, is dependent on the maximum grain size, $a_{\rm max}$ (see Figure \ref{fig:beta}). For micron-sized solids, $\beta_{\mathrm {mm}}$ does not change with $a_{\rm max}$, and is close to the average value of the interstellar medium dust; that is, ${\sim}\,1.6\,{\pm}\,0.1$ \citep{Li2001,PlanckCollaboration2016}. The vertical dashed line in Figure~\ref{fig:beta} denotes the boundary of $a_{\rm max}\,{\sim}\,0.4\,\rm{mm}$, which divides the $a_{\rm max}$ values into two regions according to the $\beta(1.3-3\,\rm{mm})$ curve. In the region on the left side, $\beta_{\mathrm {mm}}$ monotonically increases with $a_{\rm max}$, whereas $\beta_{\mathrm {mm}}$ monotonically decreases with $a_{\rm max}$ on the right side of the dashed line  \citep{Draine2006,Ricci2010a,Birnstiel2018}. The $a_{\rm max}$ boundary between the two different trends of $\beta_{\mathrm {mm}}$ depends on the wavelengths used to calculate the opacity slope. It can be seen that a $\beta$ value lower than ${\sim}\,1.7$ is a strong indication of dust grains of millimeter sizes or even larger \citep[e.g.,][]{1991ApJ...381..250B,2005ApJ...631.1134A,2007ApJ...671.1800A,Ricci2010b,2012ApJ...760L..17P,2016A&A...588A..53T,2018ApJ...859...21A}.

Constraining dust grain size from observed spectral indices is a challenging task. This is because for a $\beta_{\mathrm {mm}}$ value there may be two solutions of $a_{\rm max}$. The situation is even worse when the grain size in the disk is smaller than ${\sim}\,0.1\,{\rm mm}$, since then $\beta_{\mathrm {mm}}$ is no longer sensitive to $a_{\rm max}$ (see Figure~\ref{fig:beta}). Moreover, translating spectral indices into grain sizes requires knowledge about the dust temperature distribution in the disk, which is poorly known. In order to self-consistently calculate the dust temperature and constrain the grain size, Monte Carlo radiative transfer models are built to fit the spectral energy distribution and millimeter observations \citep[e.g.,][]{Pinte2016,2017A&A...607A..74L,2023MNRAS.518.6092L}. These sophisticated models, usually in two dimensions, can properly treat the effect of optical depth. However, this kind of methodology requires a lot of computational resources, and is therefore preferable for case studies. \citet{Sierra2019} derived an approximate solution to the radiative transfer equation for vertically isothermal disks. The emergent intensity from the disk is expressed as an analytic formula that involves the dust temperature, surface density, and grain properties, including $a_{\rm max}$. Parameter estimation based on the analytic solution is efficient and flexible, and has therefore been widely adopted in recent works \citep[e.g.,][]{Carrasco2019,Sierra2021,Macias2021,Zhang2023,Ohashi2023}. 

Nevertheless, the analytic solution needs an input for the dust temperature to obtain $\beta_{\mathrm {mm}}$ from $\alpha_{\rm mm}$, and the vertical disk structure is ignored. How, and to which degree these simplifications affect the results needs to be investigated. In this work, we aim to examine the applicability of using the analytic models to constrain the grain size, and explore under which circumstances the grain size can be properly retrieved. We first build self-consistent reference disk models with known properties, and generate synthetic millimeter continuum images. Then, we fit the synthetic images to search for the best-fit parameter set. Finally, we compare the grain sizes of the best-fit model with the reference setup to assess the quality of retrieval. The paper is organized as follows. The establishment of the reference disk models is introduced in Section~\ref{sect:reference}. Section~\ref{sect:fitting} describes the fitting process. We present and discuss the results in Section~\ref{sect:discussion}. The paper ends up with a summary in Section~\ref{sect:summary}. 

\section{Reference models}
\label{sect:reference}
We created two disk models that differ only in the radial profile of dust grain size. The model parameters are listed in Table~\ref{parameter}. We used the \texttt{RADMC-3D}\footnote{ \url{{https://www.ita.uni-heidelberg.de/~dullemond/software/radmc-3d/}}} code \citep{2012ascl.soft02015D} to calculate the dust temperature and simulate the millimeter continuum images. This section describes the model setup, including stellar parameters, density distribution, dust properties, and observational parameters. 

\begin{table}
  \centering
  \caption{Parameter values of the reference models}
  \label{parameter}
    \begin{tabular}{lc}
    \toprule
    \toprule
    Parameters & Value \\
    \midrule
    \multicolumn{2}{c}{Stellar parameters} \\
    \midrule
    $M_{\star}\ [\mathrm M_{\odot}]$    & {0.58}\\
    $T_{\mathrm {eff}}\ [\mathrm {K}]$    & {3800} \\
    $L_{\star}\ [\mathrm L_{\odot}]$    & {0.7} \\
    \midrule
    \multicolumn{2}{c}{Dust density distribution} \\
    \midrule
    $R_{\mathrm {in}}$ [AU]  & {0.1} \\
    $R_{\mathrm {out}}$ [AU]  & {100} \\
    $H_{100}$ [AU]  & {10} \\
    $\beta_{\rm {H}}$  & {1.1}  \\
    ${R_{\rm {Gap}}}$ [AU] & {20}\\
    ${\sigma_{\rm {Gap}}}$ [AU] & {10}         \\
    ${R_{\rm {Ring}}}$ [AU] & {40} \\
    ${\sigma_{\rm {Ring}}}$ [AU] & {10}         \\
    ${\eta_{\rm {Gap}}}$   & {0.99}  \\
    ${\eta_{\rm {Ring}}}$   & {3}  \\
    $M_{\rm {dust}}\ [\mathrm M_{\odot}]$    &  {$1 \times 10^{-4}$}\\
    \midrule
    \multicolumn{2}{c}{Dust properties} \\
    \midrule
	Composition    & {\texttt{dsharpopac} \citep{Birnstiel2018}} \\
    $a_{\rm min}$ [$\mu{\rm m}$]   & {0.01} \\
    \multirow{2}{*}{$a_0$ [mm]}   &  model 1: 50   \\
       &  model 2: 5  \\
    \midrule
    \multicolumn{2}{c}{Observation parameters} \\
    \midrule
    $i$ [$\circ$]  & {0}\\
    D [pc]     & {100} \\
    Beam size [{$\prime\prime$}] & {0.05}\\
    \bottomrule
    \bottomrule
    \end{tabular}%
\end{table}%

\subsection{Stellar parameters}
We assume that the disk is passively heated by stellar irradiation. The stellar properties were set according to DS Tau, which is a T Tauri star. It has a spectral type of M0.4 and an effective temperature of 3800\,K \citep{2014ApJ...786...97H}. The stellar luminosity was fixed to $0.7\,L_{\odot}$ \citep{2023MNRAS.518.6092L}. The incident stellar spectrum was taken from the Kurucz atmosphere grid, assuming solar metallicity and a $\log g\,{=}\,3.5$ \citep{1994KurCD..19.....K}. We solved the radiative transfer 
problem at 160 wavelengths that were logarithmically spaced between $0.1\,\mu{\rm m}$ and $10000\,\mu{\rm m}$.

\subsection{Dust density distribution}
\begin{figure}[htbp]
    \centering
    \includegraphics[width=\columnwidth,angle=0]{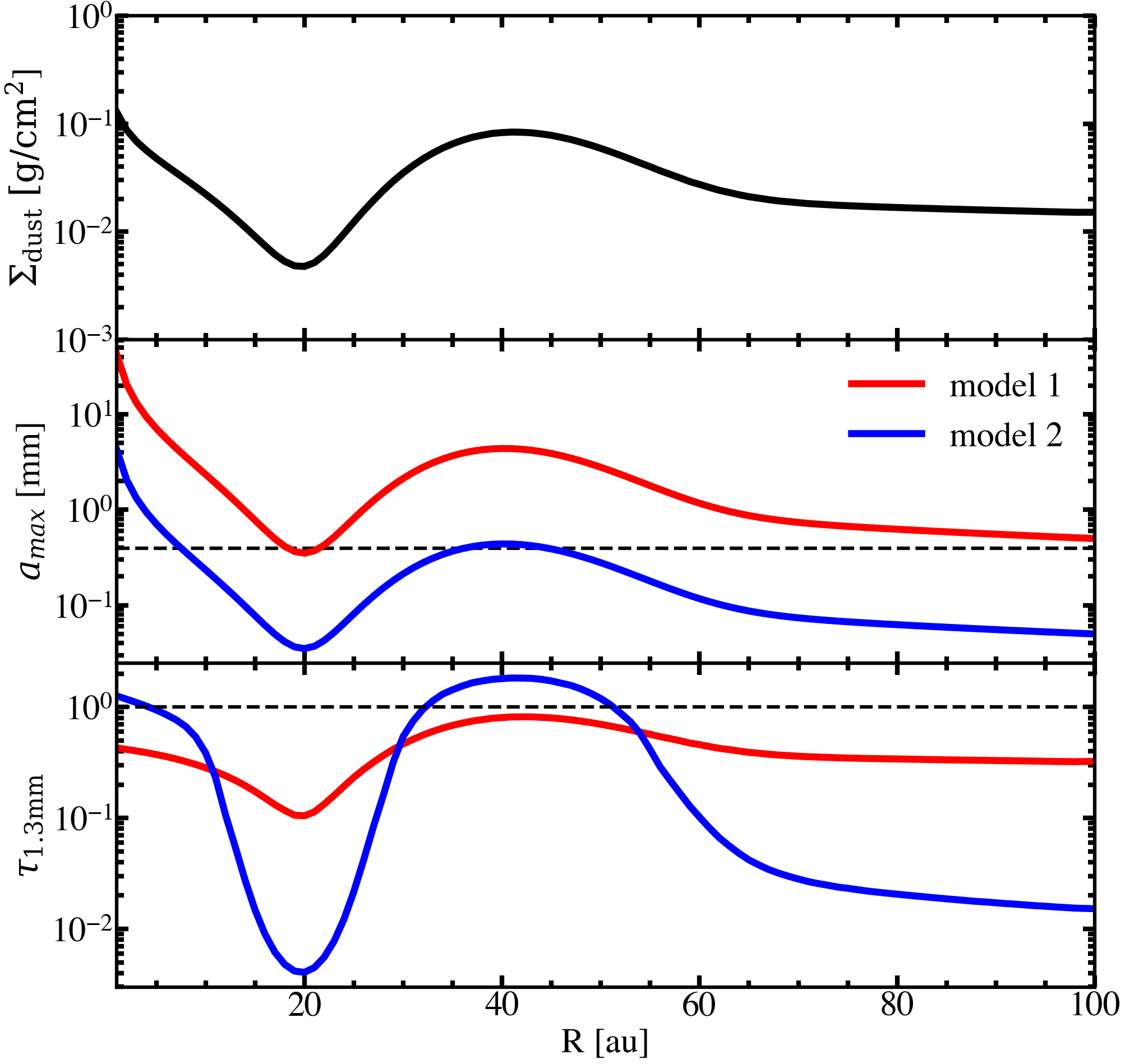}
\caption{Properties of the reference models. {\it Top panel:} Dust surface density. The dust surface densities of model 1 and model 2 are the same. {\it Middle panel:} Radial profile of the maximum grain size of model 1 (red line) and model 2 (blue line). The horizontal dashed line marks the 0.4\,mm boundary defined in Figure~\ref{fig:beta}. {\it Bottom panel:} Optical depth at the wavelength of 1.3\,mm. The color coding is identical to that in the middle panel. The $\tau_{\rm 1.3mm}\,{=}\,1$ condition is indicated with the horizontal dashed line.}
\label{fig:surdens}
\end{figure}

The disk extends from an inner radius of $R_{\rm in}\,{=}\,0.1\,\rm{AU}$ to an outer radius of $R_{\rm out}\,{=}\,100\,\rm{AU}$, and features a flared geometry. The volume density of the dust grains is given as  
\begin{equation}
\label{rho}
\rho(R, z)\,{=}\,\frac{\Sigma(R)}{\sqrt{2 \pi} H} \exp \left(-\frac{z^{2}}{2 H^{2}}\right),
\end{equation}
with the scale height
\begin{equation}
\label{h}
H\,{=}\,H_{100}\left(\frac{R}{100 \mathrm{AU}}\right)^{\beta_{\rm H}}.
\end{equation}
The slope, $\beta_{\rm H}$, describes the extent of disk flaring, and $H_{100}$ represents the scale height at the radial distance of $R\,{=}\,100\,{\rm AU}$ from the central star. We prescribed the dust surface density as 
\begin{equation}
\label{Eq:Sigma}
\begin{aligned}
\Sigma_{\rm dust}\,{=}\,\Sigma_{0} R^{-0.5} 
&\times \left\{1-{\eta_{\rm {Gap}}} \exp \left[-\frac{\left(R-{R_{\rm {Gap}}} \right)^{2}}{2 {\sigma_{\rm {Gap}}}^{2}}\right]\right\} \\
&\times \left\{1+{\eta_{\rm {Ring}}} \exp \left[-\frac{\left(R-{R_{\rm {Ring}}} \right)^{2}}{2 {\sigma_{\rm {Ring}}}^{2}}\right]\right\},
\end{aligned}
\end{equation}
where the power-law slope was set to $-0.5$, which has been found to be a reasonable value in previous works \citep{2005ApJ...627L.153D,2009ApJ...700.1502A,Williams2011,2017A&A...606A..88T,Carrasco2019}. The proportionality constant, $\Sigma_{0}$, was derived by normalizing the total dust mass, $M_{\rm dust}$, which we fixed to a typical value ($1\times10^{-4}\,M_{\odot}$) for protoplanetary disks \citep[e.g.,][]{Williams2011,Miotello2023}. To simulate a pair consisting of a gap and a ring, as is commonly observed in protoplanetary disks, a Gaussian depletion term and a Gaussian magnification term were multiplied. The position and width of the gap are denoted as ${R_{\rm {Gap}}}$ and ${\sigma_{\rm {Gap}}}$, respectively. Similarly, ${R_{\rm {Ring}}}$ and ${\sigma_{\rm {Ring}}}$ refer to the position and width of the ring. The parameter, ${\eta_{\rm {Gap}}}$, represents the depletion factor of $\Sigma_{\rm {dust}}$ in the gap, while ${\eta_{\rm {Ring}}}$ represents the magnification factor of $\Sigma_{\rm {dust}}$ in the ring. The black line in the upper panel of Figure~\ref{fig:surdens} shows the dust surface density of the reference model. We note that the dust surface densities of model 1 and model 2 are the same.

\subsection{Dust properties}
\label{section:dust}
We assume that dust grains are compact homogeneous spheres, composed of 20\% water ice, 33\% astronomical silicates, 7\% troilite, and 40\% refractory organics, with an average density of $\rho_{\mathrm {dust}}\,{=}\,1.675 \,\mathrm g/\mathrm {cm}^{-3}$. The percentages are given for the mass fraction of each component. This dust prescription was introduced to interpret the ALMA data 
from the Disk Substructures at High Angular Resolution Project \citep[DSHARP,][]{Andrews2018}. The complex refractive indices were provided by \citet{Birnstiel2018}, and the dielectric function and dust absorption or scattering coefficients were derived using the Bruggeman effective medium theory \citep{1935AnP...416..636B} and the Mie theory, respectively.

The grain size distribution is assumed to be a power law ${N(a)}\,{\propto}\,a^{-3.5}$, where $a$ is the grain radius, and ${N(a)}$ denotes the number of dust grains within the size interval $[a, a+da]$. We fixed the minimum dust size to $a_{\mathrm{min}}\,{=}\,0.01\,\mu \mathrm{m}$ and allowed the maximum grain size, $a_{\mathrm{max}}$, to vary as a function of $R$. The radial profile of $a_{\mathrm{max}}$ is similar to that of $\Sigma_{\rm dust}$, expressed as:
\begin{equation}
\label{Eq:amax}
\begin{aligned}
a_{\rm {max}}(R)\,{=}\,a_{0} R^{-1} 
&\times \left\{1-{\eta_{\rm {Gap}}} \exp \left[-\frac{\left(R-{R_{\rm {Gap}}}\right)^{2}}{2 {\sigma_{\rm {Gap}}}^{2}}\right]\right\} \\
&\times \left\{1+{\eta_{\rm {Ring}}} \exp \left[-\frac{\left(R-{R_{\rm {Ring}}}\right)^{2}}{2 {\sigma_{\rm {Ring}}}^{2}}\right]\right\},
\end{aligned}
\end{equation}
where $a_{0}$ is the $a_{\rm {max}}$ at $R\,{=}\,1\,\rm{AU}$. We set the power law index to $-1$, which has been found to be typical in protoplanetary disks \citep{2013A&A...558A..64T,2016A&A...588A..53T,2016A&A...588A.112G,Carrasco2019,Sierra2021}. The red and blue lines in the middle panel of Figure \ref{fig:surdens} show the radial profiles of $a_{\rm {max}}$ for models 1 and 2, respectively. The maximum grain size of model 1 at each radius is ten times larger than that of model 2. More importantly, all the $a_{\rm {max}}(R)$ in model 1 are larger than the $0.4\,\rm{mm}$ boundary defined in Figure \ref{fig:beta}, which implies that the model is in a parameter space with a monotonic dependence between $a_{\rm max}$ and $\beta_{\rm mm}$ if 1.3\,mm and 3.0\,mm images are used in the fitting procedure. On the contrary, $a_{\rm {max}}(R)$ in model 2 spans the $0.4\,\rm{mm}$ boundary, and hence the model is located in a parameter space with a non-monotonic dependence between $a_{\rm max}$ and $\beta_{\rm mm}$. This setup allows us to examine the performance of the retrieval framework with the analytic solution when a $\beta_{\rm mm}$ value corresponds to two $a_{\rm max}$ values (see Figure~\ref{fig:beta}). Once $a_{\rm max}$ and $\Sigma_{\rm dust}$ are known, we can calculate the optical depth, $\tau_{\nu}$. As is shown in the bottom panel of Figure~\ref{fig:surdens}, model 1 is optically thin at 1.3\,mm, while model 2 is optically thin as well, except for the very inner disk and the ring center.      

\begin{figure}[!t]
    \centering
    \includegraphics[width=\columnwidth,angle=0]{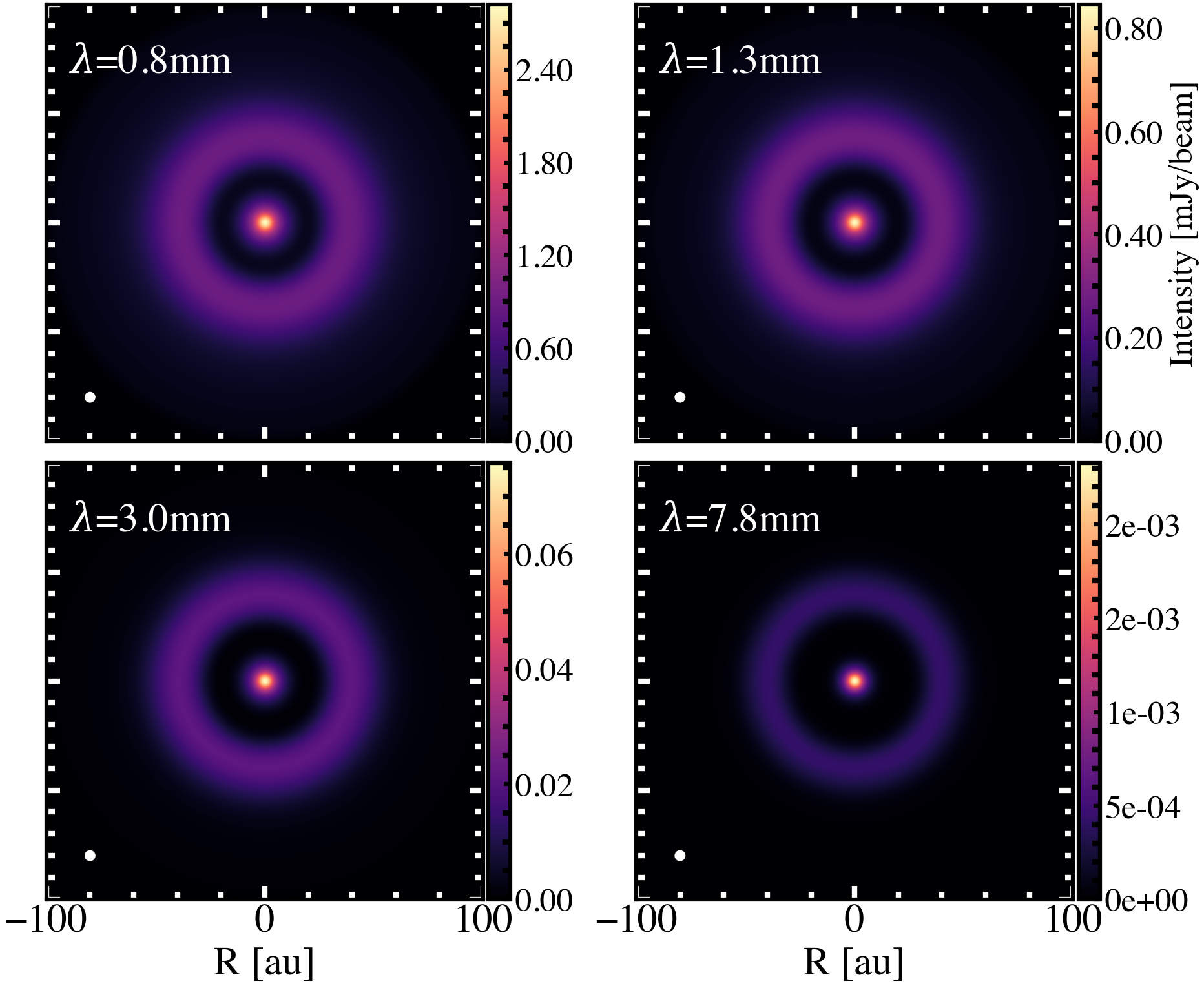}
\caption{Synthetic images of model 1 at wavelengths of 0.8\,mm, 1.3\,mm, 3.0\,mm, and 7.8\,mm. The images are observed with a face-on orientation. The beam with a size of $0.05^{\prime\prime}$ is indicated with the white dot in the lower left corner of each panel.}
\label{Fig:image}
\end{figure}

Numeric simulations of dust trapping have demonstrated that larger dust grains are more concentrated toward the ring center than the smaller ones \citep[e.g.,][] {Pinilla2012,Pinilla2015a,Pinilla2015b}. Therefore, we prescribed the radial profile of the grain size to be the same as the dust surface density, where the grain size increases in the ring and decreases in the gap. In order to have more experiments, we also ran the fitting procedure for three smooth disks without substructures. The dust surface density was fixed to $\Sigma_{\rm dust}\,{=}\,\Sigma_{0} R^{-0.5}$, and the radial profile of $a_{\rm max}$ follows a power law, $a_{\rm {max}}(R)\,{=}\,a_{0} R^{-\gamma}$, with the power law index being $\gamma\,{=}\,-0.5$, $-1$, and $-1.5$, respectively. The results of these additional tests are presented and discussed in Sect.~\ref{sect:smoothdisk}.

To set up $a_{\rm max}(R)$ in the radiative transfer simulation, we divided the disk into 40 radial bins that are linearly spaced with a width of 2.5\,AU. The value of $a_{\rm max}$ for each radial bin was determined based on the $a_{\rm max}(R)$ profile; for example, Eq.~\ref{Eq:amax}. The radial bin size was chosen to be half of the beam size (i.e., 5\,AU, see Sect.~\ref{section:obs}) of the synthetic images, ensuring that our model is sufficient to resolve the disk substructure. Using more radial bins needs more computational resources; however, the results are not significantly affected.

\subsection{Observational parameters}
\label{section:obs}
The synthetic observations were generated assuming a distance of 100\,pc. We simulated continuum images at 0.8, 1.3, 3.0, and 7.8\,mm, and convolved the raw images from \texttt{RADMC-3D} with a circular beam of $0.05^{\prime \prime}$, mimicking high-resolution ALMA and Very Large Array (VLA) observations. The beam size corresponds to a physical scale of $5\,\rm{AU}$ at the distance of 100\,pc.  The synthetic images of model 1 are shown in Figure~\ref{Fig:image}, in which the gap and ring are clearly seen. The disk inclination was set to $i\,{=}\,0^{\circ}$; a face-on orientation. We explore the effects of disk inclination on the fitting results in Sect.~\ref{sect:incl}.

We extracted the radial intensity profile from the synthetic images by applying an azimuthal average on a series of concentric ellipses with shapes determined by the disk inclination. The radial intensity profiles are the observables to be fitted.

\begin{figure}[!t]
    \centering
    \includegraphics[width=\columnwidth]{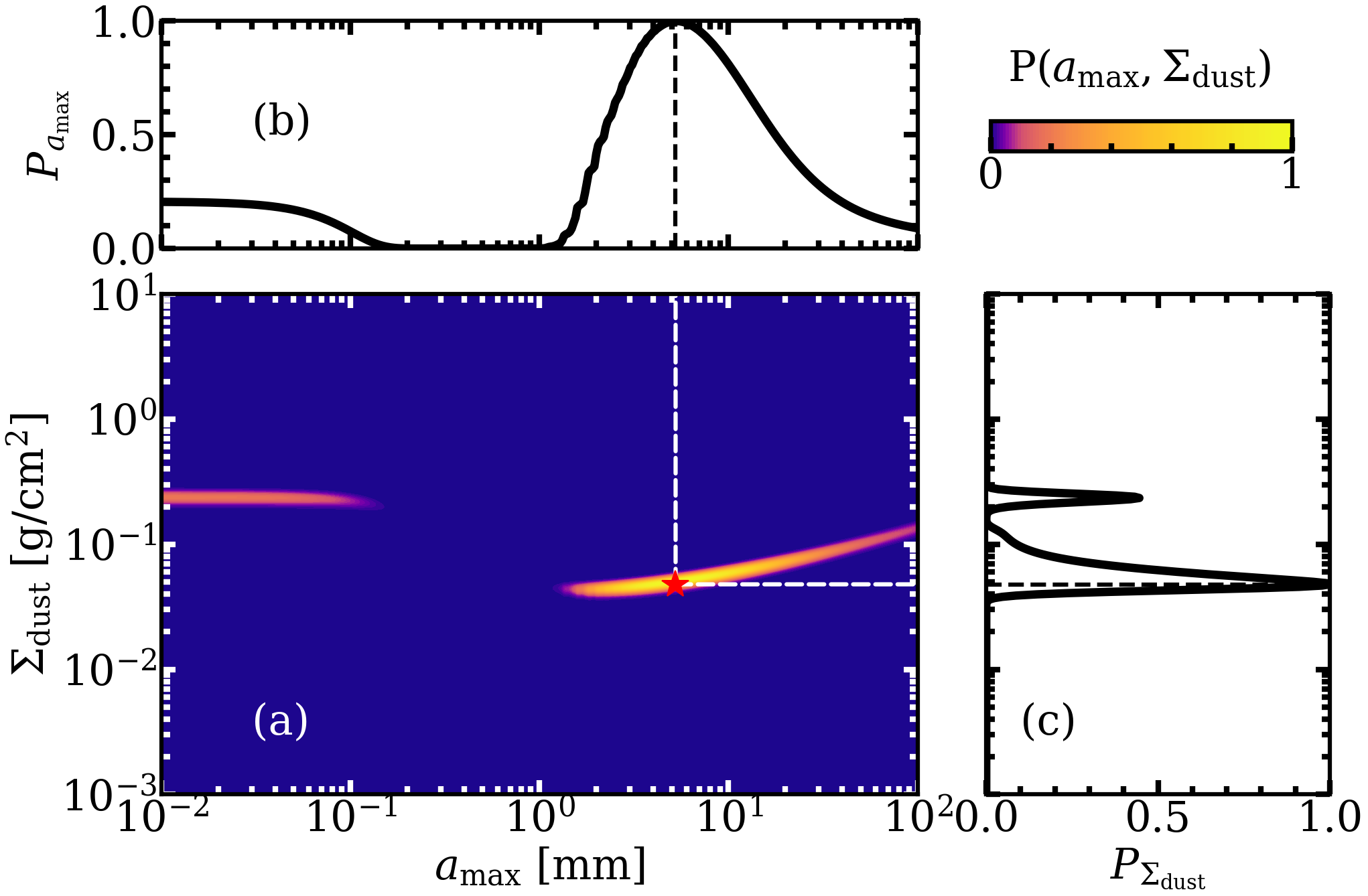}
\caption{Fitting process for model 1 at $R\,{=}\,5\,\rm{AU}$. The normalized joint probability distribution is shown as the color map in panel (a). The marginalized probabilities for $a_{\rm {max}}$ and $\Sigma_{\rm {dust}}$ are indicated with the black lines in panel (b) and (c), respectively. The red star denotes the best-fit parameter set identified from the fitting procedure.}
\label{Fig:probability}
\end{figure}

\begin{figure*}[!t]
\centering
    \includegraphics[width=0.8\textwidth]{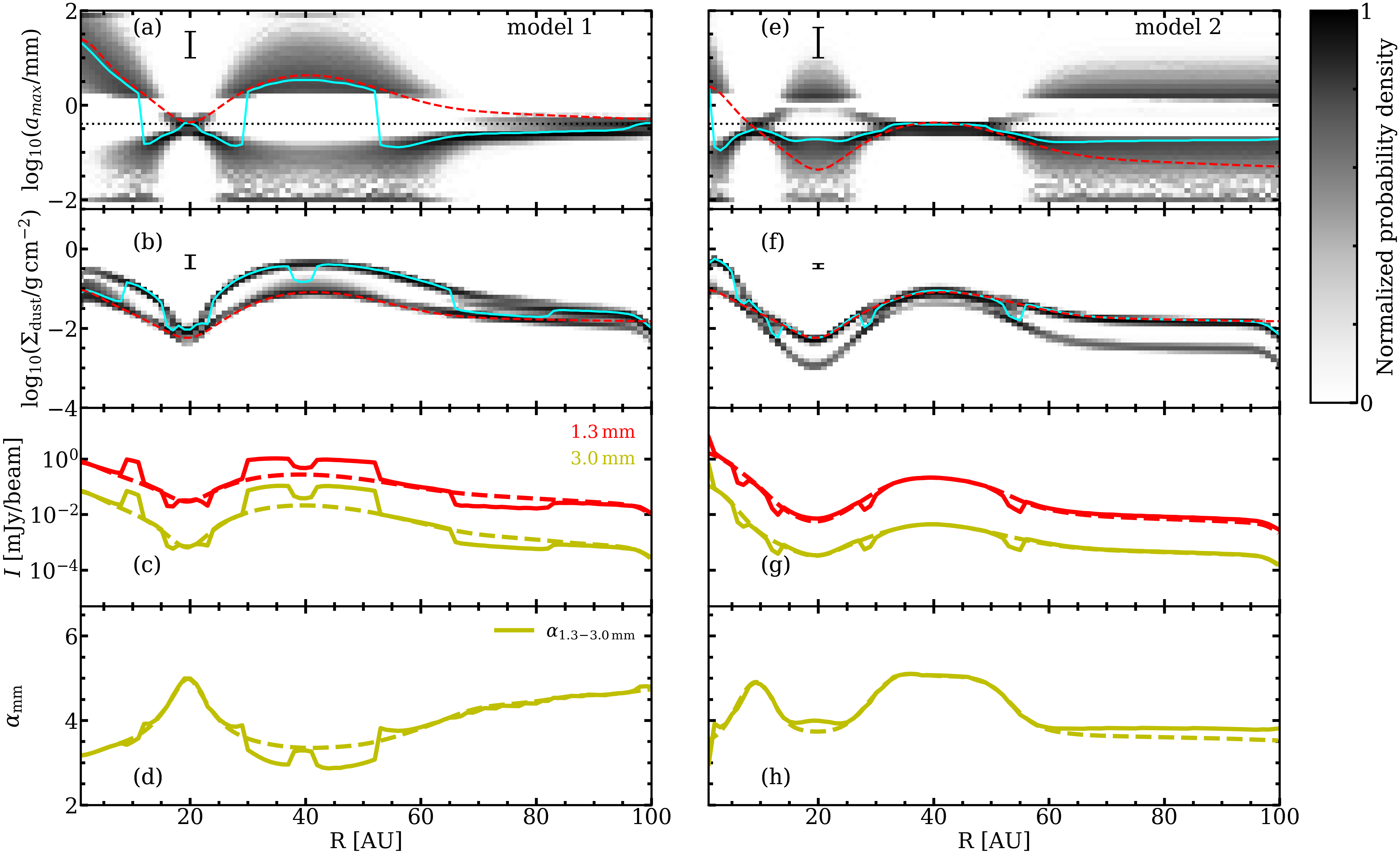}
\caption{Results of fitting to the 1.3\,mm and 3.0\,mm images for model 1 (left column) and model 2 (right column). For model 1, the best-fit $a_{\rm {max}}$ and $\Sigma_{\rm {dust}}$ are indicated with the solid cyan lines in panels (a) and (b), respectively, whereas the dashed red lines refer to the reference profiles convolved with a same beam (i.e., $0.05^{\prime\prime}$) to that of the synthetic images. The gray color scale represents the probability distribution of parameter values returned from fitting the 5000 randomly generated intensity profiles (see Sect.~\ref{sect:fitting}). The vertical bar marks the typical confidence interval of the best-fit parameter set. In panels (c) and (d), the reference profiles of intensity and spectral index are shown with dashed lines, while the best-fit results are indicated with solid lines. Panels (e)-(h) present the results for model 2.}
\label{fig:fit1}
\end{figure*}

\section{Fitting approach}
\label{sect:fitting}
To fit the synthetic images, we adopted the analytic solution to the radiative transfer equation, which was introduced by \citet{Sierra2019}. It is now a consensus that dust scattering should be considered in interpreting millimeter observations of protoplanetary disks \citep[e.g.,][]{Zhu2019,Liuh2019}. When taking dust scattering into account, the dust continuum emission from the disk can be written as 
\begin{equation}
\label{Eq:intensity}
I_\nu=B_\nu(T)\left[\left(1-\exp \left(-\tau_\nu / \mu\right)\right)+\omega_\nu F\left(\tau_\nu, \omega_\nu\right)\right],
\end{equation}
where $T$ is the dust temperature, and $\mu\,{=}\,\cos(i)$ represents the cosine of the disk inclination \citep{1993Icar..106...20M,Sierra2019}. The dust surface density ($\Sigma_{\rm {dust}}$), absorption coefficient ($\kappa_{\nu}$) and albedo ($\omega_{\nu}$) work together to determine the optical depth, $\tau_{\nu}\,{=}\,\Sigma_{\rm {dust}}\kappa_{\nu}/(1-\omega_{\nu})$. In addition, the scattering correction term ($F$) is denoted as 
\begin{equation}
\begin{aligned}
F\left(\tau_\nu, \omega_\nu\right)= & \frac{1}{\exp \left(-\sqrt{3} \epsilon_\nu \tau_\nu\right)\left(\epsilon_\nu-1\right)-\left(\epsilon_\nu+1\right)} \\
& \times\left[\frac{1-\exp \left(-\left(\sqrt{3} \epsilon_\nu+1 / \mu\right) \tau_\nu\right)}{\sqrt{3} \epsilon_\nu \mu+1}\right. \\
& \left.+\frac{\exp \left(-\tau_\nu / \mu\right)-\exp \left(-\sqrt{3} \epsilon_\nu \tau_\nu\right)}{\sqrt{3} \epsilon_\nu \mu-1}\right].
\end{aligned}
\end{equation}
The quantity $\epsilon_{\nu}$ equals $\sqrt{1-\omega_{\nu}}$. In Eq.~\ref{Eq:intensity}, we have four free parameters: $i$, $T$, $\tau_{\nu}$, and $\omega_{\nu}$. The disk inclination ($i$) can be well constrained by high-resolution observations \citep[e.g.,][]{Long2018,Huang2018}. Hence, we do not explore this parameter throughout our work. In previous studies, the dust temperature ($T$) is either considered as a free parameter \citep{Sierra2019,Macias2021,Zhang2023,Ohashi2023} or assumed to follow the expected radial profile for a passively irradiated disk \citep{Kenyon1987,Dullemond2001,Sierra2021,Doi2023}. In order to reduce the model degeneracy, we fixed $T$ at each radius to the mass-averaged mean dust temperature that was obtained from the \texttt{RADMC-3D} simulation. The optical depth ($\tau_{\nu}$) was determined by $\Sigma_{\rm {dust}}$, $\kappa_{\nu}$, and $\omega_{\nu}$. The absorption coefficient ($\kappa_{\nu}$) and albedo ($\omega_{\nu}$) depend only on the maximum grain size ($a_{\mathrm{max}}$) once the dust composition and minimum grain size ($a_{\mathrm{min}}$) are given. Therefore, we are left with a parameter space that consists of $\Sigma_{\rm {dust}}$ and $a_{\mathrm{max}}$. 

For each parameter set, \{$\Sigma_{\rm {dust}}$, $a_{\mathrm{max}}$\}, we calculated the likelihood via
\begin{equation}
\begin{aligned}
\label{Eq:likelihood}
&p\left(I_{\rm {\nu_1}},......,I_{\rm {\nu_n}} \mid a_{\max }, \Sigma_{\mathrm{dust}}\right) \propto \exp \left(-\chi^2 / 2\right),
\end{aligned}
\end{equation}
where
\begin{equation}
\begin{aligned}
&\chi^2=\sum_{n} \left(\frac{I_{\nu_n}-I_{\nu_n}^{\text {model }}}{\sigma_{\nu_n}}\right)^2
\end{aligned}
\end{equation}
is the chi-square statistic. The intensities extracted from the synthetic images at different wavelengths are denoted as $I_{\nu_1},I_{\nu_2},......,I_{\nu_n}$, while $I_{\nu_n}^{\text{model}}$ refers to the model intensity prescribed by Eq.~\ref{Eq:intensity}. The parameter $\sigma_{\nu_n}$ represents the uncertainties of the observed intensity profiles. We assume that the uncertainties are dominated by the absolute calibration error, and took 10\% for all of the four considered wavelengths. This assumption is reasonable according to ALMA and VLA observations of protoplanetary disks \citep[e.g.,][]{Long2018,Andrews2018,Francis2020,Tobin2020}. 

\begin{figure*}[!t]
\centering
    \includegraphics[width=0.8\textwidth]{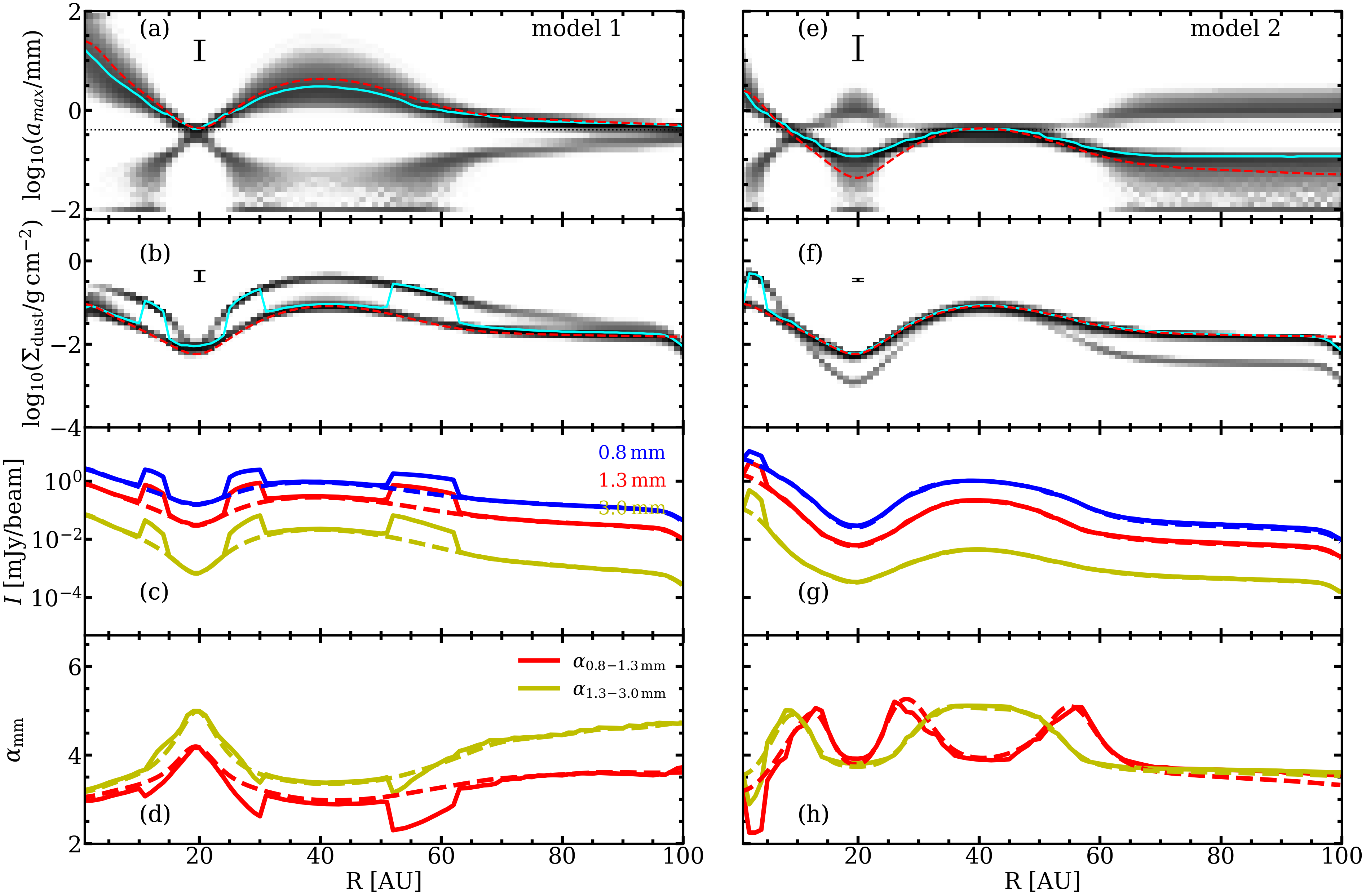}
\caption{Same as in Figure~\ref{fig:fit1}, but for simultaneously fitting to the 0.8\,mm, 1.3\,mm, and 3.0\,mm images.}
\label{fig:fit1-1}
\end{figure*}

The fitting procedure was performed at each integer value of the radius of the disk in units of AU. At each radius, we calculated a large grid of model intensities by sampling $\Sigma_{\rm {dust}}$ and $a_{\mathrm{max}}$. For $\Sigma_{\rm {dust}}$, the range for exploration is from $10^{-4}$ to $10\,\rm{g\,cm}^{-2}$. In previous works, $a_{\mathrm{max}}$ was commonly sampled at two different ranges: one is located in the small size regime, $[10\,\mu \rm {m}$, $0.4\,\rm {mm}]$, and the other is in the large size domain, $[0.4\,\rm {mm}, 10\,\rm {cm}]$ (see Figure~\ref{fig:beta}). Then the fitting was conducted twice, separately within the small and large size boundaries. The final results were obtained by combining the two fitting solutions \citep{Sierra2021,Macias2021,Zhang2023}. The purpose of this two-round practice is to reduce the degeneracy between the small and large size regimes. However, combining the results of fitting in the small and large size regimes is a challenge, and is usually guided by empirical knowledge from theories of dust evolution. For instance, it is generally believed that dust grains grow faster in regions with higher densities; for example, in the inner disk. Moreover, millimeter- and centimeter-sized dust grains would rapidly drift toward the center of the disk \citep{1977MNRAS.180...57W}. As a result, the grain population in the inner disk is dominated by large dust, while small dust grains are more concentrated within the outer disk \citep[e.g.,][]{Birnstiel2016,Birnstiel2023}. In addition, disk substructures, frequently observed with ALMA, have been interpreted as the outcome of large dust grains being trapped toward local gas pressure maxima \citep[e.g.,][] {Pinilla2012,Pinilla2015a,Pinilla2015b}. Therefore, dust grains in the rings are expected to be larger than those in the low-surface-density gaps. Nevertheless, we do not know the true $a_{\rm max}(R)$ in real protoplanetary disks. Hence, in our work, we have treated the dust size as a black box, and sampled the parameter from $10\,\mu \rm {m}$ all the way to $10\,\rm {cm}$. Both $\Sigma_{\rm {dust}}$ and $a_{\mathrm{max}}$ were sampled in a logarithmic manner with a stepwidth of 0.01 dex.

\begin{figure*}[!t]
\centering
    \includegraphics[width=0.8\textwidth]{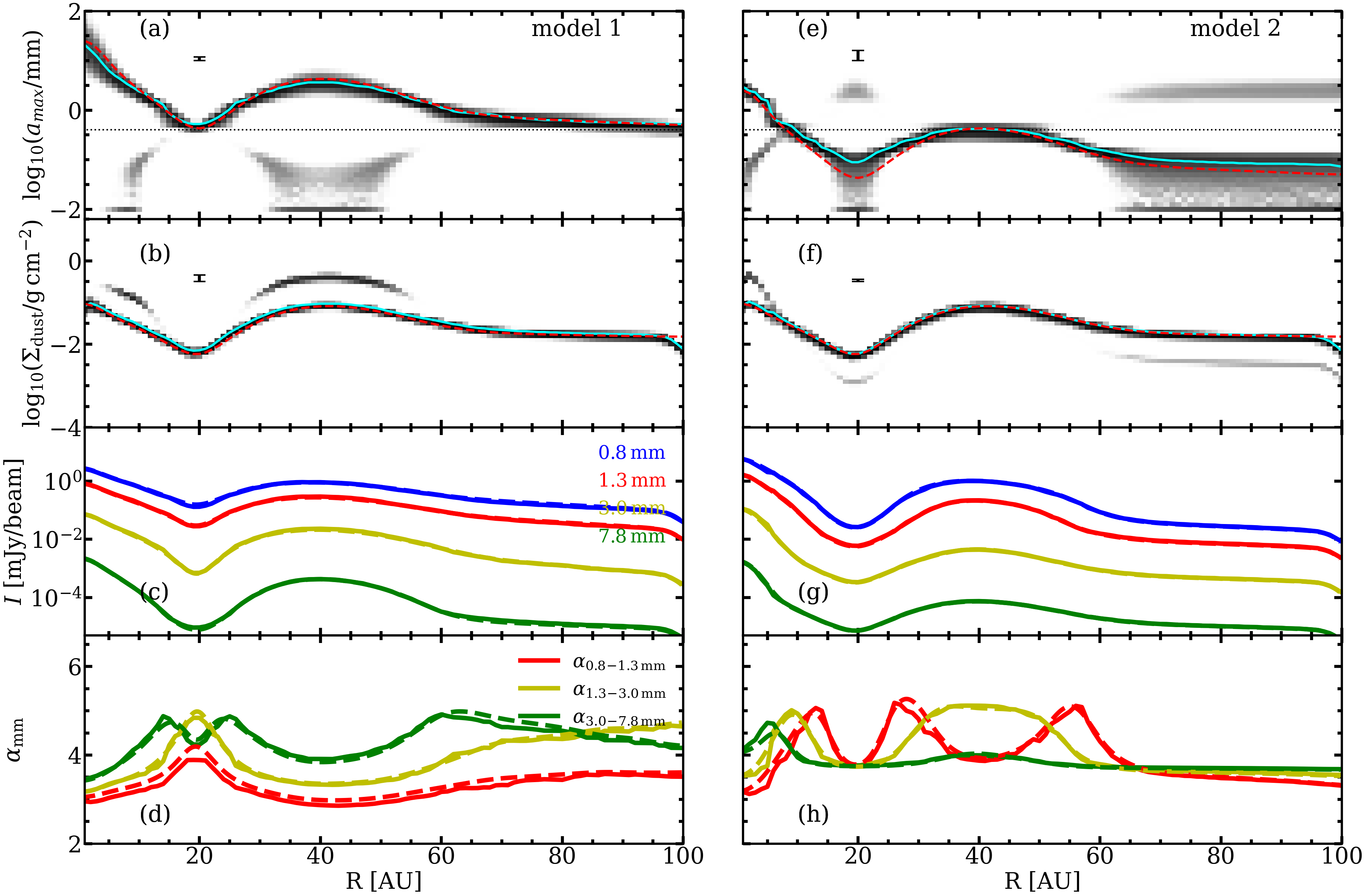}
\caption{Same as in Figure~\ref{fig:fit1}, but for simultaneously fitting to the 0.8\,mm, 1.3\,mm, 3.0\,mm, and 7.8\,mm images.}
\label{fig:fit2}
\end{figure*}

In the fitting procedure, the joint probability distribution, $P(a_{\rm {max}}, \Sigma_{\rm {dust}})$, was marginalized to obtain the probabilities of $\Sigma_{\rm {dust}}$ and $a_{\rm {max}}$. As an example, the shaded area in panel (a) of Figure~\ref{Fig:probability} illustrates the regions in the parameter space that can reproduce the observations at $R\,{=}\,5\,\rm{AU}$ in model 1. The marginalized probabilities of $a_{\rm {max}}$ and $\Sigma_{\rm {dust}}$ are shown in panels (b) and (c), respectively. The best-fit parameter set, indicated with the red star, is identified as the intersection where both $\Sigma_{\rm {dust}}$ and $a_{\rm {max}}$ reach their peak probability. Moreover, we used a Monte Carlo method to derive the final best-fit parameter set and its error in order to minimize the influence of the uncertainty of observations ($\sigma_{\nu_n}$) on the results. We generated a set of 5,000 intensity profiles at each wavelength by randomly adding Gaussian noise, with the standard deviation being 10\% of the absolute value of the intensities. The final best model was determined as the $50^{\rm {th}}$ percentile among the results from fitting the 5,000 random intensities, and the confidence interval was associated with the $25^{\rm {th}}$ and $75^{\rm {th}}$ percentiles.  The effect of the Monte Carlo method on the parameter uncertainty is described in Sect~\ref{sect:flux4}. Once the fitting procedure for each radius was done, we obtained the best-fit model profiles of $\Sigma_{\rm {dust}}$ and $a_{\rm {max}}$, and could compare with the reference setup. 

\section{Results and discussion}
\label{sect:discussion}

In this section, we present the results by fitting images at only two wavelengths, three wavelengths, and all four wavelengths. Moreover, we investigate the influence of the disk inclination on the fitting results. 

\subsection{Fitting to the 1.3 and 3.0\,mm images}
\label{sect:flux2}
The spectral index, measured between 1.3\,mm and 3.0\,mm, is suitable for grain size studies \citep[e.g.,][]{Ricci2010a,Ricci2010b,Andrews2020,Tazzari2021}. At these two wavelengths, disks have relatively strong thermal emission and low optical depth, which are favorable conditions for disk characterization given limited observation time. Optical depth is higher toward shorter wavelengths, whereas long wavelength observations take a much longer integration, and contaminations by other radiation processes, such as free-free emission, should be carefully treated \citep[e.g.,][]{Macias2021,Rota2024}. Therefore, we first fit the 1.3\,mm and 3\,mm images. The results for model 1 are presented in panels (a)-(d) of Figure~\ref{fig:fit1}. The dashed red lines in panels (a) and (b) are the reference profiles for $a_{\rm {max}}$ and $\Sigma_{\rm {dust}}$, respectively, while the cyan lines refer to the best-fit results. It should be noted that the reference model profiles were convolved with a same beam size ($0.05^{\prime\prime}$) as the observations. The gray color scale represents the probability distribution of parameter values obtained from fitting the 5000 Monte Carlo models at each radius. The vertical bar indicates the confidence interval of the best-fit result. For a better visualization, we only show the typical uncertainty that is the median of the uncertainties at all radii. A comparison of the intensities and spectral index between the reference setup and the best-fit results is given in panels (c) and (d). 

As can be seen from Figure~\ref{fig:fit1}, the best-fit $a_{\rm {max}}(R)$ based on the 1.3\,mm and 3\,mm images match with the reference values only for the inner 15\,AU and the ring region. In other parts of the disk, both $a_{\rm {max}}$ and $\Sigma_{\rm {dust}}$ were not satisfactorily retrieved. The errors are pretty large in general. For model 2, the results are similar, and are shown in panels (e)-(h) of Figure~\ref{fig:fit1}. Nevertheless, except for the inner 10\,AU, the best-fit surface densities are close to the reference setup.

The mismatch between the model and synthetic observation is mainly due to the following reasons. Images of two wavelengths can derive only one spectral index; however, there are two solutions of $a_{\rm max}$ for one particular value of spectral index even in the optically thin regime (see Figure~\ref{fig:beta}). One is located in the small size region, and the other is in the large size domain. This fact leads to an ambiguity in identifying the optimum $a_{\rm max}$, which will in turn affect the retrieval for $\Sigma_{\rm {dust}}$. Therefore, if $a_{\rm max}$ cannot be well retrieved, $\Sigma_{\rm {dust}}$ often deviates from the reference values at a significant level. Even at locations where $a_{\rm max}$ is well retrieved, $\Sigma_{\rm {dust}}$ may not be correctly obtained, directly illustrating the large uncertainty in the results based on images at only two wavelengths. 

We also examined the results by fitting other pairs of images, such as a combination of 1.3 and 7.8\,mm or 3.0 and 7.8\,mm. These attempts arrive at the similar conclusion. 

\subsection{Fitting to the 0.8, 1.3 and 3.0 images simultaneously}
\label{sect:flux3}
We tested the accuracy of the fitting by considering three images, with the results shown in Figure~\ref{fig:fit1-1}. We selected the 0.8, 1.3, and 3\,mm images as the observable, because they correspond to Band 7, Band 6, and Band 3 ALMA observations that have commonly been carried out for protoplanetary disks \citep[e.g.,][]{Pascucci2016,Andrews2018,Oberg2021}. For model 1, compared to the fitting with only two images, the retrieval for $a_{\rm max}$ in the gap and outer disk is significantly improved, and uncertainties of $a_{\rm max}$ also decrease. The dust surface densities in the ring and outer disk are better recovered. The improvements in the parameter retrieval for model 2 are less than those for model 1. 

We also considered a combination of 1.3, 3.0 and 7.8\,mm images. The results are similar to those obtained by fitting the images of the three short wavelengths, but the constraints on the dust surface density are overall better since the longest wavelength of 7.8\,mm yields the lowest optical depth.

\begin{figure*}[!t]
\centering
\includegraphics[width=0.9\textwidth]{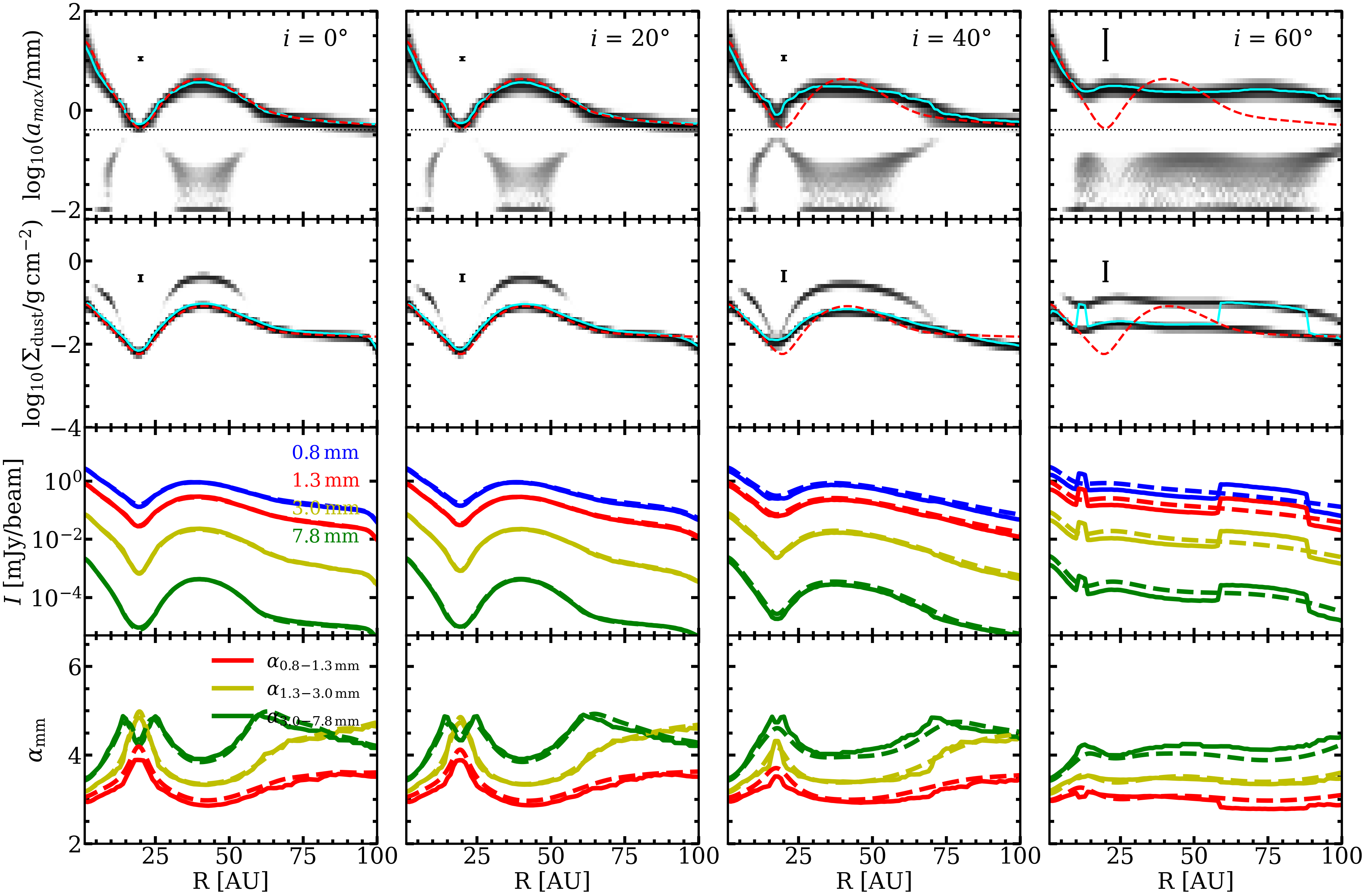}
\caption{Same as in Figure~\ref{fig:fit1}, but for investigating the effect of disk inclination on the parameter retrieval. In the fitting procedure, we adopted the setup of model 1, and took all four images into account.}
\label{fig:fit3}
\end{figure*}

\subsection{Fitting to the 0.8, 1.3, 3.0 and 7.8\,mm images simultaneously}
\label{sect:flux4}
As a further experiment, we simultaneously fit the four images. The results are presented in Figure~\ref{fig:fit2}. For model 1, the best-fit $a_{\rm {max}}(R)$ closely match the reference values at all radii. The quality of retrieval for $\Sigma_{\rm {dust}}$ is also high. For model 2, the results are also satisfactory, though the best-fit model slightly overpredicts $a_{\rm {max}}(R)$. It is evident that regardless of whether $a_{\rm {max}}$ crosses the boundary between the small and large size regions, fitting images at four wavelengths constrain the dust grain size well. This can be understood because the number of data points is larger than that of the free parameters. One can derive only one spectral index with two wavelengths. Three spectral indices can be obtained by using three images. When four images are considered, there will be six different combinations. Figure~\ref{fig:beta} shows three spectral indices as an illustration, $\beta(0.8-1.3\,\rm{mm})$, $\beta(1.3-3\,\rm{mm})$, and $\beta(3-7.8\,\rm{mm})$, which vary differently with $a_{\rm max}$. This highly reduces the model degeneracy.

In the inner disk (i.e., $R\,{\sim}\,10\,\rm{AU}$) and the ring, the best-fit $a_{\rm max}$ and $\Sigma_{\rm dust}$ of the 5,000 Monte Carlo models still span a large range of values, similar to the outcome of model 1 when fitting with two images (see Figure~\ref{fig:fit1}).
Our work shows that the Monte Carlo approach described in Sect.~\ref{sect:fitting} helps to obtain more accurate grain sizes and dust surface densities, because fitting to the observable only once probably returns $a_{\rm {max}}$ and $\Sigma_{\rm {dust}}$ profiles that fluctuate significantly with radius (see Figure 12 in \citet{Sierra2021} and Figure 4 in \citet{Zhang2023}, for instance). The strong fluctuation is induced by the model degeneracy. Especially, when the observed spectral index, $\alpha_{\rm mm}$, corresponds to an opacity slope, $\beta_{\rm mm}$, close to ${\sim}\,1.8$ (see Figure~\ref{fig:beta}), there exists two solutions; that is, the low-surface-density large size solution and high-surface-density small size solution (see panel (a) of Figure~\ref{Fig:probability}). With the Monte Carlo approach, we fit the intensities that were generated by randomly adding Gaussian noise to the reference intensity profiles, and obtained a result that favors the reference setup.


\begin{figure*}[!h]
\centering
    \includegraphics[width=0.9\textwidth]{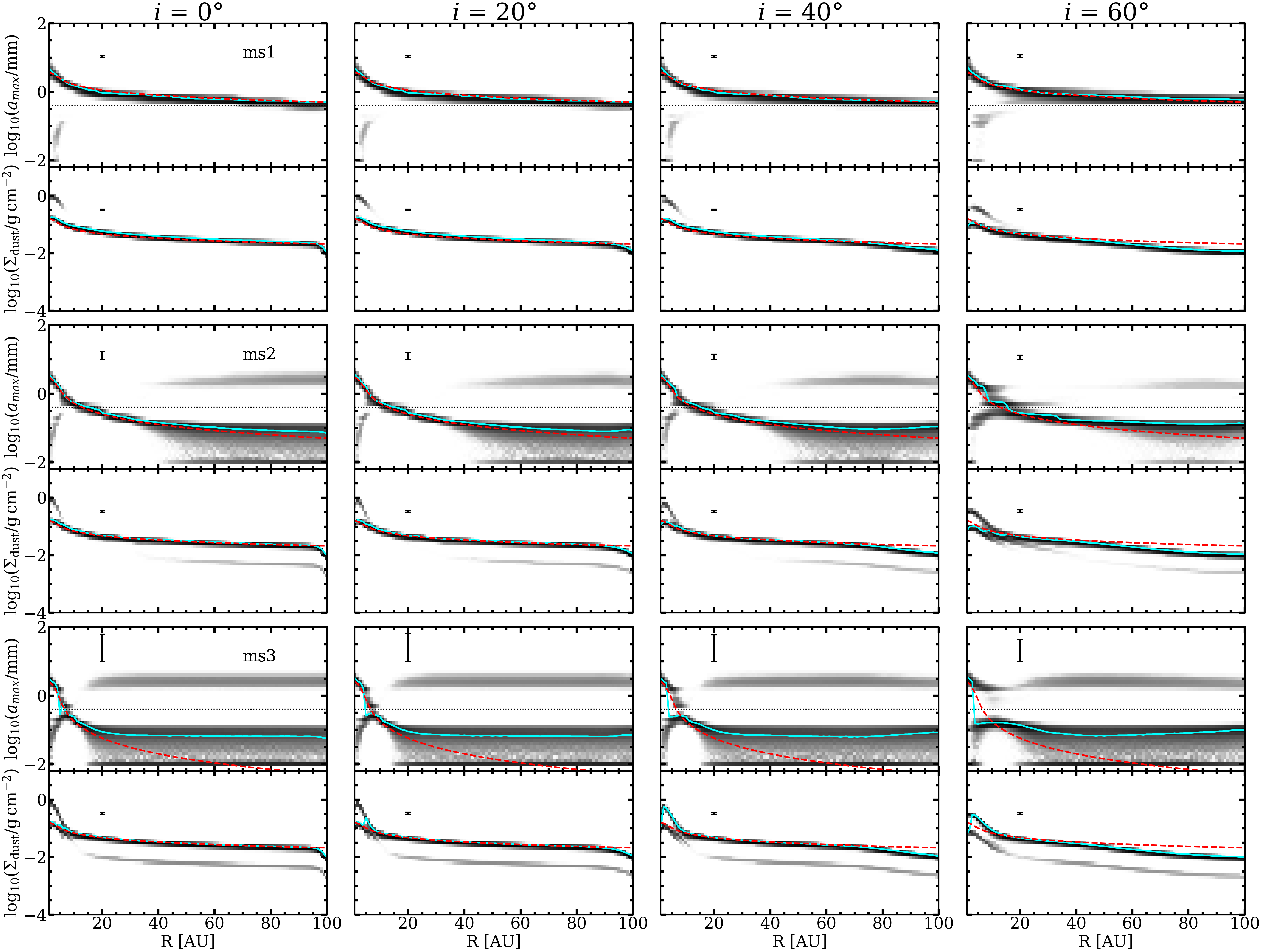}
\caption{Same as in panels (a) and (b) of Figure~\ref{fig:fit1}; however, the reference setup does not include substructures, mimicking smooth disks. The three reference models, denoted as ms1, ms2, and ms3, differ in terms of the radial grain size profile (see Sect.~\ref{sect:smoothdisk}).}
\label{fig:fit4}
\end{figure*}

\subsection{Effects of the disk inclination}
\label{sect:incl}
The above experiments were all conducted for face-on disks. In this section, we investigate the effect of disk inclination on the fitting results by tilting model 1 with different angles. Figure~\ref{fig:fit3} displays the results for the inclination of $20^\circ$, $40^\circ$, and $60^\circ$, respectively. When the disk inclination is low — for example, ${\lesssim}\,20^\circ$ — the fitting quality is minimally affected, and both $a_{\rm {max}}$ and $\Sigma_{\rm {dust}}$ can be well retrieved. As the inclination increases to $40^{\circ}$, the best-fit profile for both parameters starts to deviate from the reference profile, with the gap-ring contrast being shallower than the reference. When the disk is highly inclined — for example, $i\,{=}\,60^{\circ}$ — the retrieved results are not reliable for most parts of the disk, though radial intensity profiles can still be well recovered at all wavelengths.  

Constraining $a_{\rm {max}}$ and $\Sigma_{\rm {dust}}$ for disks with high inclinations is challenging. The observed intensity is an integration of dust emission along the line of sight. Therefore, substructures will be smoothed out when the disk is tilted, leading to smooth intensity profiles rather than showing distinct gap and ring features. Accordingly, the retrieved profiles for $a_{\rm {max}}$ and $\Sigma_{\rm {dust}}$ are also flat. For face-on disks, the constrained $a_{\rm {max}}$ and $\Sigma_{\rm {dust}}$ trace the mean disk properties over a radius range determined by the beam size; $5\,\rm{AU}$ in our case. However, for inclined disks, the radius range for averaging is larger due to the inclination effect, making the parameter retrieval more complex and difficult. Therefore, special attention should be paid in the investigation of dust grain size in disks with high inclinations. 

\begin{table}[!t]
\centering
  \caption{Parameter of reference models for smooth disks}
  \label{table:amaxb}
    \begin{tabular}{ccc}
    \toprule
    \toprule
    ID     & $a_{0}$ [mm] & $\gamma$ \\
    \midrule
    ms1    & 5  & 0.5 \\
    ms2    & 5  & 1.0 \\
    ms3    & 5  & 1.5 \\
    \bottomrule
    \bottomrule
    \end{tabular}%
\end{table}%

\subsection{Fitting to reference models of smooth disks}
\label{sect:smoothdisk}
{The reference models used for the above tests have a gap and a ring (see Eq.~\ref{Eq:Sigma} and \ref{Eq:amax}). We also examined the retrieval process using three smooth disks without substructures. For this purpose, we prescribed the dust surface density as $\Sigma(R)\,{=}\,\Sigma_{0}\,R^{-0.5}$. The three disks differ in terms of the radial grain size profile given by $a_{\rm {max}}(R)\,{=}\,a_{0}\,R^{-\gamma}$. Table~\ref{table:amaxb} tabulates the model parameters. For each of the three reference disks, we fit the 0.8, 1.3, 3.0, and 7.8\,mm images simultaneously. The results are presented in Figure~\ref{fig:fit4}.}

These additional experiments demonstrate again that the analytic approach recovers the grain size well when four images are considered in the fitting process. As the disk inclination increases, the best-fit results appear inconsistent with the reference setup, particularly for disks with a large $\gamma$. Since $a_{0}$ is fixed, a steeper grain size profile means that more disk regions are populated with $a_{\rm max}\,{<}\,0.1\,{\rm mm}$. Figure~\ref{fig:fit4} shows that it is difficult to recover grain sizes smaller than 0.1\,mm. The reason is that all of the three $\beta_{\mathrm{mm}}$ are close to ${\sim}\,1.8$, and no longer sensitive to the grain size in this regime (see Figure~\ref{fig:beta}).

\section{Summary}
\label{sect:summary}

Investigating the grain size and its correlation with disk substructures is crucial to understand the initial step of planet formation. Constraining the grain size from millimeter observations needs to self-consistently solve the radiative transfer equation, which is time-consuming and not efficient for parameter studies with large sample sizes. As an alternative, analytic models under the assumption of vertically isothermal disks have been proposed and widely used in the literature.     

In this work, we created reference disk models with known properties, and simulated millimeter continuum images as synthetic observations. We fit the synthetic images using analytic models to investigate the applicability of the approach, and examined the conditions under which the input grain sizes can be recovered. Our conclusions are given as follows:

\begin{itemize}
  \item[(1)] Fitting images at only two wavelengths is not sufficient to retrieve the reference setup, mainly due to the degeneracy between the small and large size solutions (see Figure~\ref{fig:beta}). Fitting three images improves the retrieval of grain size; however, the obtained dust surface density still differs a lot from the reference model. The analytic models work well when all of the four images are considered in the fitting process. By combining each pair of images from the four, one can derive six spectral indices that behave differently as a function of $a_{\rm max}$ (see three of the six spectral indices in Figure~\ref{fig:beta}). This highly reduces the model degeneracy.  
    
  \item[(2)] The inclination angle has a significant impact on the fitting results. When the inclination is low — for example, ${\lesssim}\,40^{\circ}$ — the reference grain sizes can be recovered well using the analytic models. However, when the disk is highly tilted, both the grain size and the dust surface density cannot be constrained. This is because the inclination effect will smooth out any substructure in the radial intensity profile of the disk, and the optical depth effect needs to be treated more properly. 
  
  \item[(3)] The uncertainties of the best-fit parameter set are relatively large in some regions; for example, in the inner disk and the ring. The Monte Carlo method introduced in Sect.~\ref{sect:fitting} helps to obtain more accurate results. 
\end{itemize}

\begin{acknowledgements}
We thank the anonymous referee for the constructive comments that improve our manuscript. YL acknowledges financial supports by the National Natural Science Foundation of China (Grant Number 11973090), and by the International Partnership Program of Chinese Academy of Sciences (Grant Number 019GJHZ2023016FN).  HW acknowledges the financial support by the National Natural Science Foundation of China (Grant Number 11973091).
\end{acknowledgements}

\bibliographystyle{aa}
\bibliography{grainsize.bib}

\begin{thebibliography}{76}
\expandafter\ifx\csname natexlab\endcsname\relax\def\natexlab#1{#1}\fi

\bibitem[{{Adachi} {et~al.}(1976){Adachi}, {Hayashi}, \&
  {Nakazawa}}]{1976PThPh..56.1756A}
{Adachi}, I., {Hayashi}, C., \& {Nakazawa}, K. 1976, Progress of Theoretical
  Physics, 56, 1756

\bibitem[{{Andrews}(2020)}]{Andrews2020}
{Andrews}, S.~M. 2020, \araa, 58, 483

\bibitem[{{Andrews} {et~al.}(2018){Andrews}, {Huang}, {P{\'e}rez}, {Isella},
  {Dullemond}, {Kurtovic}, {Guzm{\'a}n}, {Carpenter}, {Wilner}, {Zhang}, {Zhu},
  {Birnstiel}, {Bai}, {Benisty}, {Hughes}, {{\"O}berg}, \&
  {Ricci}}]{Andrews2018}
{Andrews}, S.~M., {Huang}, J., {P{\'e}rez}, L.~M., {et~al.} 2018, \apjl, 869,
  L41

\bibitem[{{Andrews} \& {Williams}(2005)}]{2005ApJ...631.1134A}
{Andrews}, S.~M. \& {Williams}, J.~P. 2005, \apj, 631, 1134

\bibitem[{{Andrews} \& {Williams}(2007)}]{2007ApJ...671.1800A}
{Andrews}, S.~M. \& {Williams}, J.~P. 2007, \apj, 671, 1800

\bibitem[{{Andrews} {et~al.}(2009){Andrews}, {Wilner}, {Hughes}, {Qi}, \&
  {Dullemond}}]{2009ApJ...700.1502A}
{Andrews}, S.~M., {Wilner}, D.~J., {Hughes}, A.~M., {Qi}, C., \& {Dullemond},
  C.~P. 2009, \apj, 700, 1502

\bibitem[{{Ansdell} {et~al.}(2018){Ansdell}, {Williams}, {Trapman}, {van
  Terwisga}, {Facchini}, {Manara}, {van der Marel}, {Miotello}, {Tazzari},
  {Hogerheijde}, {Guidi}, {Testi}, \& {van Dishoeck}}]{2018ApJ...859...21A}
{Ansdell}, M., {Williams}, J.~P., {Trapman}, L., {et~al.} 2018, \apj, 859, 21

\bibitem[{{Armitage}(2010)}]{Armitage2010}
{Armitage}, P.~J. 2010, {Astrophysics of Planet Formation}

\bibitem[{{Bae} {et~al.}(2023){Bae}, {Isella}, {Zhu}, {Martin}, {Okuzumi}, \&
  {Suriano}}]{Bae2023}
{Bae}, J., {Isella}, A., {Zhu}, Z., {et~al.} 2023, in Astronomical Society of
  the Pacific Conference Series, Vol. 534, Protostars and Planets VII, ed.
  S.~{Inutsuka}, Y.~{Aikawa}, T.~{Muto}, K.~{Tomida}, \& M.~{Tamura}, 423

\bibitem[{{Beckwith} \& {Sargent}(1991)}]{1991ApJ...381..250B}
{Beckwith}, S. V.~W. \& {Sargent}, A.~I. 1991, \apj, 381, 250

\bibitem[{{Birnstiel}(2023)}]{Birnstiel2023}
{Birnstiel}, T. 2023, arXiv e-prints, arXiv:2312.13287

\bibitem[{{Birnstiel} {et~al.}(2009){Birnstiel}, {Dullemond}, \&
  {Brauer}}]{2009A&A...503L...5B}
{Birnstiel}, T., {Dullemond}, C.~P., \& {Brauer}, F. 2009, \aap, 503, L5

\bibitem[{{Birnstiel} {et~al.}(2018){Birnstiel}, {Dullemond}, {Zhu}, {Andrews},
  {Bai}, {Wilner}, {Carpenter}, {Huang}, {Isella}, {Benisty}, {P{\'e}rez}, \&
  {Zhang}}]{Birnstiel2018}
{Birnstiel}, T., {Dullemond}, C.~P., {Zhu}, Z., {et~al.} 2018, \apjl, 869, L45

\bibitem[{{Birnstiel} {et~al.}(2016){Birnstiel}, {Fang}, \&
  {Johansen}}]{Birnstiel2016}
{Birnstiel}, T., {Fang}, M., \& {Johansen}, A. 2016, \ssr, 205, 41

\bibitem[{{Brauer} {et~al.}(2008){Brauer}, {Dullemond}, \&
  {Henning}}]{2008A&A...480..859B}
{Brauer}, F., {Dullemond}, C.~P., \& {Henning}, T. 2008, \aap, 480, 859

\bibitem[{{Brauer} {et~al.}(2007){Brauer}, {Dullemond}, {Johansen}, {Henning},
  {Klahr}, \& {Natta}}]{Brauer2007}
{Brauer}, F., {Dullemond}, C.~P., {Johansen}, A., {et~al.} 2007, \aap, 469,
  1169

\bibitem[{{Bruggeman}(1935)}]{1935AnP...416..636B}
{Bruggeman}, D.~A.~G. 1935, Annalen der Physik, 416, 636

\bibitem[{{Carrasco-Gonz{\'a}lez} {et~al.}(2019){Carrasco-Gonz{\'a}lez},
  {Sierra}, {Flock}, {Zhu}, {Henning}, {Chandler}, {Galv{\'a}n-Madrid},
  {Mac{\'\i}as}, {Anglada}, {Linz}, {Osorio}, {Rodr{\'\i}guez}, {Testi},
  {Torrelles}, {P{\'e}rez}, \& {Liu}}]{Carrasco2019}
{Carrasco-Gonz{\'a}lez}, C., {Sierra}, A., {Flock}, M., {et~al.} 2019, \apj,
  883, 71

\bibitem[{{Chokshi} {et~al.}(1993){Chokshi}, {Tielens}, \&
  {Hollenbach}}]{Chokshi1993}
{Chokshi}, A., {Tielens}, A.~G.~G.~M., \& {Hollenbach}, D. 1993, \apj, 407, 806

\bibitem[{{Davis}(2005)}]{2005ApJ...627L.153D}
{Davis}, S.~S. 2005, \apjl, 627, L153

\bibitem[{{Doi} \& {Kataoka}(2023)}]{Doi2023}
{Doi}, K. \& {Kataoka}, A. 2023, \apj, 957, 11

\bibitem[{{Dominik} \& {Tielens}(1997)}]{Dominik1997}
{Dominik}, C. \& {Tielens}, A.~G.~G.~M. 1997, \apj, 480, 647

\bibitem[{{Draine}(2006)}]{Draine2006}
{Draine}, B.~T. 2006, \apj, 636, 1114

\bibitem[{{Dr{\k{a}}{\.z}kowska} {et~al.}(2023){Dr{\k{a}}{\.z}kowska},
  {Bitsch}, {Lambrechts}, {Mulders}, {Harsono}, {Vazan}, {Liu}, {Ormel},
  {Kretke}, \& {Morbidelli}}]{Drazkowska2023}
{Dr{\k{a}}{\.z}kowska}, J., {Bitsch}, B., {Lambrechts}, M., {et~al.} 2023, in
  Astronomical Society of the Pacific Conference Series, Vol. 534, Protostars
  and Planets VII, ed. S.~{Inutsuka}, Y.~{Aikawa}, T.~{Muto}, K.~{Tomida}, \&
  M.~{Tamura}, 717

\bibitem[{{Dullemond} {et~al.}(2001){Dullemond}, {Dominik}, \&
  {Natta}}]{Dullemond2001}
{Dullemond}, C.~P., {Dominik}, C., \& {Natta}, A. 2001, \apj, 560, 957

\bibitem[{{Dullemond} {et~al.}(2012){Dullemond}, {Juhasz}, {Pohl}, {Sereshti},
  {Shetty}, {Peters}, {Commercon}, \& {Flock}}]{2012ascl.soft02015D}
{Dullemond}, C.~P., {Juhasz}, A., {Pohl}, A., {et~al.} 2012, {RADMC-3D: A
  multi-purpose radiative transfer tool}

\bibitem[{{Francis} {et~al.}(2020){Francis}, {Johnstone}, {Herczeg}, {Hunter},
  \& {Harsono}}]{Francis2020}
{Francis}, L., {Johnstone}, D., {Herczeg}, G., {Hunter}, T.~R., \& {Harsono},
  D. 2020, \aj, 160, 270

\bibitem[{{Guidi} {et~al.}(2016){Guidi}, {Tazzari}, {Testi}, {de
  Gregorio-Monsalvo}, {Chandler}, {P{\'e}rez}, {Isella}, {Natta}, {Ortolani},
  {Henning}, {Corder}, {Linz}, {Andrews}, {Wilner}, {Ricci}, {Carpenter},
  {Sargent}, {Mundy}, {Storm}, {Calvet}, {Dullemond}, {Greaves}, {Lazio},
  {Deller}, \& {Kwon}}]{2016A&A...588A.112G}
{Guidi}, G., {Tazzari}, M., {Testi}, L., {et~al.} 2016, \aap, 588, A112

\bibitem[{{Herczeg} \& {Hillenbrand}(2014)}]{2014ApJ...786...97H}
{Herczeg}, G.~J. \& {Hillenbrand}, L.~A. 2014, \apj, 786, 97

\bibitem[{{Huang} {et~al.}(2018){Huang}, {Andrews}, {Dullemond}, {Isella},
  {P{\'e}rez}, {Guzm{\'a}n}, {{\"O}berg}, {Zhu}, {Zhang}, {Bai}, {Benisty},
  {Birnstiel}, {Carpenter}, {Hughes}, {Ricci}, {Weaver}, \&
  {Wilner}}]{Huang2018}
{Huang}, J., {Andrews}, S.~M., {Dullemond}, C.~P., {et~al.} 2018, \apjl, 869,
  L42

\bibitem[{{Kenyon} \& {Hartmann}(1987)}]{Kenyon1987}
{Kenyon}, S.~J. \& {Hartmann}, L. 1987, \apj, 323, 714

\bibitem[{{Kley} \& {Nelson}(2012)}]{Kley2012}
{Kley}, W. \& {Nelson}, R.~P. 2012, \araa, 50, 211

\bibitem[{{Kurucz}(1994)}]{1994KurCD..19.....K}
{Kurucz}, R. 1994, Solar abundance model atmospheres for 0, 19

\bibitem[{{Lesur} {et~al.}(2022){Lesur}, {Ercolano}, {Flock}, {Lin}, {Yang},
  {Barranco}, {Benitez-Llambay}, {Goodman}, {Johansen}, {Klahr}, {Laibe},
  {Lyra}, {Marcus}, {Nelson}, {Squire}, {Simon}, {Turner}, {Umurhan}, \&
  {Youdin}}]{Lesur2022}
{Lesur}, G., {Ercolano}, B., {Flock}, M., {et~al.} 2022, arXiv e-prints,
  arXiv:2203.09821

\bibitem[{{Li} \& {Draine}(2001)}]{Li2001}
{Li}, A. \& {Draine}, B.~T. 2001, \apj, 554, 778

\bibitem[{{Li} {et~al.}(2023){Li}, {Liu}, {Wang}, {Wang}, \&
  {Ma}}]{2023MNRAS.518.6092L}
{Li}, D., {Liu}, Y., {Wang}, H., {Wang}, Y., \& {Ma}, Y. 2023, \mnras, 518,
  6092

\bibitem[{{Liu}(2019)}]{Liuh2019}
{Liu}, H.~B. 2019, \apjl, 877, L22

\bibitem[{{Liu} {et~al.}(2017){Liu}, {Henning}, {Carrasco-Gonz{\'a}lez},
  {Chandler}, {Linz}, {Birnstiel}, {van Boekel}, {P{\'e}rez}, {Flock}, {Testi},
  {Rodr{\'\i}guez}, \& {Galv{\'a}n-Madrid}}]{2017A&A...607A..74L}
{Liu}, Y., {Henning}, T., {Carrasco-Gonz{\'a}lez}, C., {et~al.} 2017, \aap,
  607, A74

\bibitem[{{Long} {et~al.}(2018){Long}, {Pinilla}, {Herczeg}, {Harsono},
  {Dipierro}, {Pascucci}, {Hendler}, {Tazzari}, {Ragusa}, {Salyk}, {Edwards},
  {Lodato}, {van de Plas}, {Johnstone}, {Liu}, {Boehler}, {Cabrit}, {Manara},
  {Menard}, {Mulders}, {Nisini}, {Fischer}, {Rigliaco}, {Banzatti}, {Avenhaus},
  \& {Gully-Santiago}}]{Long2018}
{Long}, F., {Pinilla}, P., {Herczeg}, G.~J., {et~al.} 2018, \apj, 869, 17

\bibitem[{{Mac{\'\i}as} {et~al.}(2021){Mac{\'\i}as}, {Guerra-Alvarado},
  {Carrasco-Gonz{\'a}lez}, {Ribas}, {Espaillat}, {Huang}, \&
  {Andrews}}]{Macias2021}
{Mac{\'\i}as}, E., {Guerra-Alvarado}, O., {Carrasco-Gonz{\'a}lez}, C., {et~al.}
  2021, \aap, 648, A33

\bibitem[{{Miotello} {et~al.}(2023){Miotello}, {Kamp}, {Birnstiel}, {Cleeves},
  \& {Kataoka}}]{Miotello2023}
{Miotello}, A., {Kamp}, I., {Birnstiel}, T., {Cleeves}, L.~C., \& {Kataoka}, A.
  2023, in Astronomical Society of the Pacific Conference Series, Vol. 534,
  Protostars and Planets VII, ed. S.~{Inutsuka}, Y.~{Aikawa}, T.~{Muto},
  K.~{Tomida}, \& M.~{Tamura}, 501

\bibitem[{{Miyake} \& {Nakagawa}(1993)}]{1993Icar..106...20M}
{Miyake}, K. \& {Nakagawa}, Y. 1993, \icarus, 106, 20

\bibitem[{{Nakagawa} {et~al.}(1986){Nakagawa}, {Sekiya}, \&
  {Hayashi}}]{1986Icar...67..375N}
{Nakagawa}, Y., {Sekiya}, M., \& {Hayashi}, C. 1986, \icarus, 67, 375

\bibitem[{{{\"O}berg} {et~al.}(2021){{\"O}berg}, {Guzm{\'a}n}, {Walsh},
  {Aikawa}, {Bergin}, {Law}, {Loomis}, {Alarc{\'o}n}, {Andrews}, {Bae},
  {Bergner}, {Boehler}, {Booth}, {Bosman}, {Calahan}, {Cataldi}, {Cleeves},
  {Czekala}, {Furuya}, {Huang}, {Ilee}, {Kurtovic}, {Le Gal}, {Liu}, {Long},
  {M{\'e}nard}, {Nomura}, {P{\'e}rez}, {Qi}, {Schwarz}, {Sierra}, {Teague},
  {Tsukagoshi}, {Yamato}, {van't Hoff}, {Waggoner}, {Wilner}, \&
  {Zhang}}]{Oberg2021}
{{\"O}berg}, K.~I., {Guzm{\'a}n}, V.~V., {Walsh}, C., {et~al.} 2021, \apjs,
  257, 1

\bibitem[{{Ohashi} {et~al.}(2023){Ohashi}, {Momose}, {Kataoka}, {Higuchi},
  {Tsukagoshi}, {Ueda}, {Codella}, {Podio}, {Hanawa}, {Sakai}, {Kobayashi},
  {Okuzumi}, \& {Tanaka}}]{Ohashi2023}
{Ohashi}, S., {Momose}, M., {Kataoka}, A., {et~al.} 2023, \apj, 954, 110

\bibitem[{{Paardekooper} {et~al.}(2022){Paardekooper}, {Dong}, {Duffell},
  {Fung}, {Masset}, {Ogilvie}, \& {Tanaka}}]{Paardekooper2022}
{Paardekooper}, S.-J., {Dong}, R., {Duffell}, P., {et~al.} 2022, arXiv
  e-prints, arXiv:2203.09595

\bibitem[{{Pascucci} {et~al.}(2016){Pascucci}, {Testi}, {Herczeg}, {Long},
  {Manara}, {Hendler}, {Mulders}, {Krijt}, {Ciesla}, {Henning}, {Mohanty},
  {Drabek-Maunder}, {Apai}, {Sz{\H{u}}cs}, {Sacco}, \&
  {Olofsson}}]{Pascucci2016}
{Pascucci}, I., {Testi}, L., {Herczeg}, G.~J., {et~al.} 2016, \apj, 831, 125

\bibitem[{{P{\'e}rez} {et~al.}(2012){P{\'e}rez}, {Carpenter}, {Chandler},
  {Isella}, {Andrews}, {Ricci}, {Calvet}, {Corder}, {Deller}, {Dullemond},
  {Greaves}, {Harris}, {Henning}, {Kwon}, {Lazio}, {Linz}, {Mundy}, {Sargent},
  {Storm}, {Testi}, \& {Wilner}}]{2012ApJ...760L..17P}
{P{\'e}rez}, L.~M., {Carpenter}, J.~M., {Chandler}, C.~J., {et~al.} 2012,
  \apjl, 760, L17

\bibitem[{{Pinilla} {et~al.}(2012{\natexlab{a}}){Pinilla}, {Benisty}, \&
  {Birnstiel}}]{Pinilla2012}
{Pinilla}, P., {Benisty}, M., \& {Birnstiel}, T. 2012{\natexlab{a}}, \aap, 545,
  A81

\bibitem[{{Pinilla} {et~al.}(2012{\natexlab{b}}){Pinilla}, {Birnstiel},
  {Ricci}, {Dullemond}, {Uribe}, {Testi}, \& {Natta}}]{2012A&A...538A.114P}
{Pinilla}, P., {Birnstiel}, T., {Ricci}, L., {et~al.} 2012{\natexlab{b}}, \aap,
  538, A114

\bibitem[{{Pinilla} {et~al.}(2015{\natexlab{a}}){Pinilla}, {de Juan Ovelar},
  {Ataiee}, {Benisty}, {Birnstiel}, {van Dishoeck}, \& {Min}}]{Pinilla2015a}
{Pinilla}, P., {de Juan Ovelar}, M., {Ataiee}, S., {et~al.} 2015{\natexlab{a}},
  \aap, 573, A9

\bibitem[{{Pinilla} {et~al.}(2015{\natexlab{b}}){Pinilla}, {van der Marel},
  {P{\'e}rez}, {van Dishoeck}, {Andrews}, {Birnstiel}, {Herczeg},
  {Pontoppidan}, \& {van Kempen}}]{Pinilla2015b}
{Pinilla}, P., {van der Marel}, N., {P{\'e}rez}, L.~M., {et~al.}
  2015{\natexlab{b}}, \aap, 584, A16

\bibitem[{{Pinte} {et~al.}(2016){Pinte}, {Dent}, {M{\'e}nard}, {Hales}, {Hill},
  {Cortes}, \& {de Gregorio-Monsalvo}}]{Pinte2016}
{Pinte}, C., {Dent}, W.~R.~F., {M{\'e}nard}, F., {et~al.} 2016, \apj, 816, 25

\bibitem[{{Planck Collaboration} {et~al.}(2016){Planck Collaboration},
  {Aghanim}, {Ashdown}, {Aumont}, {Baccigalupi}, {Ballardini}, {Banday},
  {Barreiro}, {Bartolo}, {Basak}, {Benabed}, {Bernard}, {Bersanelli},
  {Bielewicz}, {Bonavera}, {Bond}, {Borrill}, {Bouchet}, {Boulanger},
  {Burigana}, {Calabrese}, {Cardoso}, {Carron}, {Chiang}, {Colombo}, {Comis},
  {Couchot}, {Coulais}, {Crill}, {Curto}, {Cuttaia}, {de Bernardis}, {de
  Zotti}, {Delabrouille}, {Di Valentino}, {Dickinson}, {Diego}, {Dor{\'e}},
  {Douspis}, {Ducout}, {Dupac}, {Dusini}, {Elsner}, {En{\ss}lin}, {Eriksen},
  {Falgarone}, {Fantaye}, {Finelli}, {Forastieri}, {Frailis}, {Fraisse},
  {Franceschi}, {Frolov}, {Galeotta}, {Galli}, {Ganga}, {G{\'e}nova-Santos},
  {Gerbino}, {Ghosh}, {Giraud-H{\'e}raud}, {Gonz{\'a}lez-Nuevo}, {G{\'o}rski},
  {Gruppuso}, {Gudmundsson}, {Hansen}, {Helou}, {Henrot-Versill{\'e}},
  {Herranz}, {Hivon}, {Huang}, {Jaffe}, {Jones}, {Keih{\"a}nen}, {Keskitalo},
  {Kiiveri}, {Kisner}, {Krachmalnicoff}, {Kunz}, {Kurki-Suonio}, {Lamarre},
  {Langer}, {Lasenby}, {Lattanzi}, {Lawrence}, {Le Jeune}, {Levrier}, {Lilje},
  {Lilley}, {Lindholm}, {L{\'o}pez-Caniego}, {Ma}, {Mac{\'\i}as-P{\'e}rez},
  {Maggio}, {Maino}, {Mandolesi}, {Mangilli}, {Maris}, {Martin},
  {Mart{\'\i}nez-Gonz{\'a}lez}, {Matarrese}, {Mauri}, {McEwen}, {Melchiorri},
  {Mennella}, {Migliaccio}, {Miville-Desch{\^e}nes}, {Molinari}, {Moneti},
  {Montier}, {Morgante}, {Moss}, {Natoli}, {Oxborrow}, {Pagano}, {Paoletti},
  {Patanchon}, {Perdereau}, {Perotto}, {Pettorino}, {Piacentini},
  {Plaszczynski}, {Polastri}, {Polenta}, {Puget}, {Rachen}, {Racine},
  {Reinecke}, {Remazeilles}, {Renzi}, {Rocha}, {Rosset}, {Rossetti}, {Roudier},
  {Rubi{\~n}o-Mart{\'\i}n}, {Ruiz-Granados}, {Salvati}, {Sandri}, {Savelainen},
  {Scott}, {Sirignano}, {Sirri}, {Soler}, {Spencer}, {Suur-Uski}, {Tauber},
  {Tavagnacco}, {Tenti}, {Toffolatti}, {Tomasi}, {Tristram}, {Trombetti},
  {Valiviita}, {Van Tent}, {Vielva}, {Villa}, {Vittorio}, {Wandelt}, {Wehus},
  {Zacchei}, \& {Zonca}}]{PlanckCollaboration2016}
{Planck Collaboration}, {Aghanim}, N., {Ashdown}, M., {et~al.} 2016, Astronomy
  and Astrophysics, 596, A109

\bibitem[{{Ricci} {et~al.}(2010{\natexlab{a}}){Ricci}, {Testi}, {Natta}, \&
  {Brooks}}]{Ricci2010b}
{Ricci}, L., {Testi}, L., {Natta}, A., \& {Brooks}, K.~J. 2010{\natexlab{a}},
  \aap, 521, A66

\bibitem[{{Ricci} {et~al.}(2010{\natexlab{b}}){Ricci}, {Testi}, {Natta},
  {Neri}, {Cabrit}, \& {Herczeg}}]{Ricci2010a}
{Ricci}, L., {Testi}, L., {Natta}, A., {et~al.} 2010{\natexlab{b}}, \aap, 512,
  A15

\bibitem[{{Rosotti} {et~al.}(2020){Rosotti}, {Teague}, {Dullemond}, {Booth}, \&
  {Clarke}}]{2020MNRAS.495..173R}
{Rosotti}, G.~P., {Teague}, R., {Dullemond}, C., {Booth}, R.~A., \& {Clarke},
  C.~J. 2020, \mnras, 495, 173

\bibitem[{{Rota} {et~al.}(2024){Rota}, {Meijerhof}, {van der Marel}, {Francis},
  {van der Tak}, \& {Sellek}}]{Rota2024}
{Rota}, A.~A., {Meijerhof}, J.~D., {van der Marel}, N., {et~al.} 2024, arXiv
  e-prints, arXiv:2401.05798

\bibitem[{{Sierra} {et~al.}(2019){Sierra}, {Lizano}, {Mac{\'\i}as},
  {Carrasco-Gonz{\'a}lez}, {Osorio}, \& {Flock}}]{Sierra2019}
{Sierra}, A., {Lizano}, S., {Mac{\'\i}as}, E., {et~al.} 2019, \apj, 876, 7

\bibitem[{{Sierra} {et~al.}(2021){Sierra}, {P{\'e}rez}, {Zhang}, {Law},
  {Guzm{\'a}n}, {Qi}, {Bosman}, {{\"O}berg}, {Andrews}, {Long}, {Teague},
  {Booth}, {Walsh}, {Wilner}, {M{\'e}nard}, {Cataldi}, {Czekala}, {Bae},
  {Huang}, {Bergner}, {Ilee}, {Benisty}, {Le Gal}, {Loomis}, {Tsukagoshi},
  {Liu}, {Yamato}, \& {Aikawa}}]{Sierra2021}
{Sierra}, A., {P{\'e}rez}, L.~M., {Zhang}, K., {et~al.} 2021, \apjs, 257, 14

\bibitem[{{Takeuchi} \& {Lin}(2002)}]{Takeuchi2002}
{Takeuchi}, T. \& {Lin}, D.~N.~C. 2002, \apj, 581, 1344

\bibitem[{{Tazzari} {et~al.}(2016){Tazzari}, {Testi}, {Ercolano}, {Natta},
  {Isella}, {Chandler}, {P{\'e}rez}, {Andrews}, {Wilner}, {Ricci}, {Henning},
  {Linz}, {Kwon}, {Corder}, {Dullemond}, {Carpenter}, {Sargent}, {Mundy},
  {Storm}, {Calvet}, {Greaves}, {Lazio}, \& {Deller}}]{2016A&A...588A..53T}
{Tazzari}, M., {Testi}, L., {Ercolano}, B., {et~al.} 2016, \aap, 588, A53

\bibitem[{{Tazzari} {et~al.}(2017){Tazzari}, {Testi}, {Natta}, {Ansdell},
  {Carpenter}, {Guidi}, {Hogerheijde}, {Manara}, {Miotello}, {van der Marel},
  {van Dishoeck}, \& {Williams}}]{2017A&A...606A..88T}
{Tazzari}, M., {Testi}, L., {Natta}, A., {et~al.} 2017, \aap, 606, A88

\bibitem[{{Tazzari} {et~al.}(2021){Tazzari}, {Testi}, {Natta}, {Williams},
  {Ansdell}, {Carpenter}, {Facchini}, {Guidi}, {Hogherheijde}, {Manara},
  {Miotello}, \& {van der Marel}}]{Tazzari2021}
{Tazzari}, M., {Testi}, L., {Natta}, A., {et~al.} 2021, \mnras, 506, 5117

\bibitem[{{Teague} {et~al.}(2018{\natexlab{a}}){Teague}, {Bae}, {Bergin},
  {Birnstiel}, \& {Foreman-Mackey}}]{2018ApJ...860L..12T}
{Teague}, R., {Bae}, J., {Bergin}, E.~A., {Birnstiel}, T., \& {Foreman-Mackey},
  D. 2018{\natexlab{a}}, \apjl, 860, L12

\bibitem[{{Teague} {et~al.}(2018{\natexlab{b}}){Teague}, {Bae}, {Birnstiel}, \&
  {Bergin}}]{2018ApJ...868..113T}
{Teague}, R., {Bae}, J., {Birnstiel}, T., \& {Bergin}, E.~A.
  2018{\natexlab{b}}, \apj, 868, 113

\bibitem[{{Testi} {et~al.}(2014){Testi}, {Birnstiel}, {Ricci}, {Andrews},
  {Blum}, {Carpenter}, {Dominik}, {Isella}, {Natta}, {Williams}, \&
  {Wilner}}]{Testi2014}
{Testi}, L., {Birnstiel}, T., {Ricci}, L., {et~al.} 2014, in Protostars and
  Planets VI, ed. H.~{Beuther}, R.~S. {Klessen}, C.~P. {Dullemond}, \&
  T.~{Henning}, 339

\bibitem[{{Tobin} {et~al.}(2020){Tobin}, {Sheehan}, {Megeath},
  {D{\'\i}az-Rodr{\'\i}guez}, {Offner}, {Murillo}, {van 't Hoff}, {van
  Dishoeck}, {Osorio}, {Anglada}, {Furlan}, {Stutz}, {Reynolds}, {Karnath},
  {Fischer}, {Persson}, {Looney}, {Li}, {Stephens}, {Chandler}, {Cox},
  {Dunham}, {Tychoniec}, {Kama}, {Kratter}, {Kounkel}, {Mazur}, {Maud},
  {Patel}, {Perez}, {Sadavoy}, {Segura-Cox}, {Sharma}, {Stephenson}, {Watson},
  \& {Wyrowski}}]{Tobin2020}
{Tobin}, J.~J., {Sheehan}, P.~D., {Megeath}, S.~T., {et~al.} 2020, \apj, 890,
  130

\bibitem[{{Trotta} {et~al.}(2013){Trotta}, {Testi}, {Natta}, {Isella}, \&
  {Ricci}}]{2013A&A...558A..64T}
{Trotta}, F., {Testi}, L., {Natta}, A., {Isella}, A., \& {Ricci}, L. 2013,
  \aap, 558, A64

\bibitem[{{Weidenschilling}(1977)}]{1977MNRAS.180...57W}
{Weidenschilling}, S.~J. 1977, \mnras, 180, 57

\bibitem[{{Whipple}(1972)}]{1972fpp..conf..211W}
{Whipple}, F.~L. 1972, in From Plasma to Planet, ed. A.~{Elvius}, 211

\bibitem[{{Williams} \& {Cieza}(2011)}]{Williams2011}
{Williams}, J.~P. \& {Cieza}, L.~A. 2011, \araa, 49, 67

\bibitem[{{Youdin} \& {Shu}(2002)}]{Youdin2002}
{Youdin}, A.~N. \& {Shu}, F.~H. 2002, \apj, 580, 494

\bibitem[{{Zhang} {et~al.}(2015){Zhang}, {Blake}, \& {Bergin}}]{Zhang2015}
{Zhang}, K., {Blake}, G.~A., \& {Bergin}, E.~A. 2015, \apjl, 806, L7

\bibitem[{{Zhang} {et~al.}(2023){Zhang}, {Zhu}, {Ueda}, {Kataoka}, {Sierra},
  {Carrasco-Gonz{\'a}lez}, \& {Mac{\'\i}as}}]{Zhang2023}
{Zhang}, S., {Zhu}, Z., {Ueda}, T., {et~al.} 2023, \apj, 953, 96

\bibitem[{{Zhu} {et~al.}(2019){Zhu}, {Zhang}, {Jiang}, {Kataoka}, {Birnstiel},
  {Dullemond}, {Andrews}, {Huang}, {P{\'e}rez}, {Carpenter}, {Bai}, {Wilner},
  \& {Ricci}}]{Zhu2019}
{Zhu}, Z., {Zhang}, S., {Jiang}, Y.-F., {et~al.} 2019, \apjl, 877, L18

\end{thebibliography}

\end{document}